\documentclass{aa}  

\usepackage{graphicx}
\usepackage{txfonts,textcomp}
\usepackage[hidelinks,pdftex]{hyperref}
\begin{document} 

\title{Forgotten treasures in the HST/FOC UV imaging polarimetric archives of active galactic nuclei}

\subtitle{I. Pipeline and benchmarking against NGC~1068 and exploring IC~5063}

\author{T. Barnouin\inst{1}\thanks{\href{mailto:thibault.barnouin@astro.unistra.fr}{thibault.barnouin@astro.unistra.fr}}
    \and
    F. Marin\inst{1}
    \and
    E. Lopez-Rodriguez\inst{2}
    \and
    L. Huber\inst{1}          
    \and
    M. Kishimoto\inst{3}
}

\institute{Universit\'e de Strasbourg, CNRS, Observatoire astronomique de Strasbourg, UMR 7550, F-67000 Strasbourg, France
    \and
    Kavli Institute for Particle Astrophysics and Cosmology (KIPAC), Stanford University, USA
    \and
    Department of Astrophysics \& Atmospheric Sciences, Kyoto Sangyo University, Kamigamo-motoyama, Kita-ku, Kyoto 603-8555, Japan
}

\date{Received July 3, 2023; accepted Month Day, 2023}

\abstract
{Over its 13 years of operation (1990 -- 2002), the Faint Object Camera (FOC)  on board the Hubble Space Telescope (HST) observed 26 individual active galactic nuclei (AGNs) in ultraviolet (UV) imaging polarimetry. However, not all of the observations have been reduced and analyzed or set within a standardized framework.}
{We plan to reduce and analyze the AGN observations that have been neglected in the FOC archives using a consistent, novel, and open-access reduction pipeline of our own. We then extend the method to the full AGN sample, thus leading to potential discoveries in the near future.}
{We developed a new pipeline in Python that will be able to reduce all the FOC observations in imaging polarimetry in a homogeneous way. Most of the previously published reduced observations are dispersed throughout the literature, with the range of different analyses and approaches making it difficult to fully interpret the FOC AGN sample. By standardizing the method, we have enabled a coherent comparison among the different observational sets.}
{In this first paper of a series  exploring the full HST/FOC AGN sample, we present an exhaustively detailed account of how to properly reduce the observational data. Current progress in cross-correlation functions, convolution kernels, and a sophisticated merging and smoothing of the various polarization filter images, together with precise propagation of errors, has provided state-of-the-art UV polarimetric maps. We compare our new maps to the benchmark AGN case of NGC~1068 and successfully reproduce the main results previously published, while pushing the polarimetric exploration of this AGN futher, thanks to  a finer resolution and a higher signal-to-noise ratio (S/N) than previously reported. We also present, for the first time, an optical polarimetric map of the radio-loud AGN IC~5063 and we examine the complex interactions between the AGN outflows and the surrounding interstellar medium (ISM).}
{Thanks to our newly and standardized reduction pipeline, we were able to explore the full HST/FOC AGN sample, starting with observations that had not been previously published  (e.g., IC~5063 here). This pipeline will allow us to make a complete atlas of UV polarimetric images of the 26 unique AGNs observed by the FOC, highlighting the importance and necessity of (imaging) polarimeters for the upcoming new generation of 30-m class telescopes.}


\maketitle
\section{Introduction}

Polarimetry has proven to be one of the most resourceful observational methods in astronomy \citep{Hildebrand2005,Hough2006}. From stars to planets, supernovae remnants to gamma-ray bursts, polarimetry has brought a wealth of information about the geometry and composition of cosmic sources, including magnetic field intensity and topology in both small- and large-scale structures. However, it is probably the field of active galactic nuclei (AGNs) that polarimetry has contributed the most \citep[and references therein]{Marin2019}, starting with the proposition of a unified model for AGNs from optical and ultraviolet (UV) polarimetry \citep{Antonucci1985}, followed by the uncovering of a near-infrared (NIR) polarized accretion disk spectrum in quasars \citep{Kishimoto2008}, or, most recently,  the dichotomy of radio-loud and radio-quiet quasars in far-infrared (FIR) polarimetry \citep{Lopez-Rodriguez2022}.

One of the major challenges with AGNs is that they are extra-galactic, compact objects that cannot be resolved using conventional imaging techniques.  As the typical scale of the internal disk of an AGN is $\sim$2pc, for NGC~1068 (most studied Type 2 AGN at $\sim$13.5Mpc), it would require a spatial resolution of less than 10 milliarcseconds to resolve. Even interferometry is only able to resolve a couple of the closest or most massive sources \citep[e.g.,][]{Akiyama2019,GR2022,Isbell2022}. This method recently was able to get to milliarcseconds resolution with VLTI/MIDI and VLTI/MATISSE in the mid-IR and 100 microarcseconds resolution with CHARA array in  the near-IR that has allowed for the  resolution the nuclei of NGC~1068 \citep{Lopez-Gonzaga2014,GR2022}, Circinus galaxy \citep{Tristram2014,Isbell2022}, and NGC4151 \citep{Kishimoto2022}. Polarimetry has the unique advantage of transcending spatial constraints: polarimetric techniques can provide physical information well below the beam of the observations because only the polarized source is measured, while the unpolarized light does not contribute to the total polarized light. A disk's morphology can be easily told apart from a spherical morphology, even if the region is beyond the resolving power of the best imaging telescope. Using this pivotal characteristic, it is strikingly evident why polarimetry is the key to improving our understanding of the compact and luminous regions at the center of host galaxies known as AGNs. Their characteristics indicate that their intrinsic luminosity is not produced by stars. Instead, a supermassive black hole \citep[10$^5$ -- 10$^{10}$ solar masses,][]{Shankar2004,Inayoshi2016} is thought to accrete matter at the center of the AGN, releasing thermal radiation that peaks in the blue-UV band of the electromagnetic spectrum \citep{Shakura1973}. Surrounding this sub-parsec-scale central engine, there are outflows, relativistic jets, clouds of gas with various ionization stages, and a reservoir of dust and molecular gas and stars. The global picture of AGNs is certainly not easy to decipher and even the most up-to-date models struggle with establishing their true locations, compositions, and generation mechanisms of their broad variety of structures. The most enigmatic of all structures lies in the accreting region surrounding the central supermassive black hole \citep{Antonucci1993,Urry1995,Elvis2000,Netzer2015}. This is an area where polarimetry can prove especially handy. In addition, since the accretion engine emits the most in the UV band, where the starlight polluting contribution is weak, it can be most efficient to observe AGNs in this specific waveband. 

Today, there are no longer any far- or mid-UV polarimeters available for AGN observations. A few telescopes mounted with spectropolarimetric instruments reaching the near-UV band still exist (such as the VLT/FORS2 or the HST/WFPC2) but it would be necessary to observe high-redshift sources in order to probe the far-UV band\footnote{For a galaxy at redshift $z$ = 3, the Lyman break will appear to be at wavelengths of about 3600~\AA}. This is highly limiting since high-redshift objects are often fainter than AGNs from the nearby Universe and the required amount of time to reach a minimal signal-to-noise ratio (S/N$\ge$ 3) in polarization is unfeasible. To examine the UV polarization of AGNs, we must rely on past instruments, namely the Wisconsin Ultraviolet Photopolarimeter Experiment \citep[WUPPE,][]{Nordsieck1982,Stanford1985,Code1989} and various instruments aboard the Hubble Space Telescope (HST). WUPPE provided the first UV polarized spectra of 5 AGNs \citep[NGC~4151, NGC~1068, 3C~273, Centaurus~A, and Mrk~421; see][]{Marin2018}. On the other hand, a total of 2,000 datasets were acquired by the various UV polarimeters that equipped the HST from Cycle 0 to Cycle 22. This corresponds to several dozens of AGNs. Among those observations, for each of these instruments, a fraction of the observations  lack any associated publication. As an example, 19\% of AGN proposals in the FOC archives are yet to be published, implying a deep potential pool of scientific discoveries. 

The FOC served as a particularly interesting polarimeter. It was one of HST's five instruments at launch and consisted of a long-focal-ratio, photon-counting imager capable of taking high-resolution images in the wavelength range 1150 -- 6500~\AA. When corrected by COSTAR, the field-of-view (FoV) and pixel size of the f/96 camera were 7" $\times$ 7" (512 $\times$ 512 pixel$^2$ format) and 0.014" $\times$ 0.014", respectively. Other configurations were also possible thanks to a different optical relay (f/48). The excellent spatial resolution offered by the FOC, coupled to the very low instrumental polarization and excellent polarizing efficiencies of the polarizers in the f/96 relay made the FOC a unique instrument. No other polarimeter (or any non-polarimetric instruments for that matter) have fully used the spatial resolution capabilities of the Optical Telescope Assembly (OTA) of the HST. Unfortunately, the FOC was replaced by the ACS during Servicing Mission 3B (March 7, 2002). 

Because the FOC was ahead of its time and was one of the most promising instruments to achieve great discoveries in the field of AGNs, it is a pity that 19\% of the AGN proposals in the FOC archives lack any exploitation (5/26 AGNs observed with the FOC were never published) . Therefore, in this series of papers, we have decided to propose a rigorous, systematically complete, and consistent re-analysis of all raw HST imaging polarimetric AGN observations from the FOC in the HST archive to enable science deferred or unachieved by many approved programs.

In 2005, the Canadian Astronomy Data Center (CADC), in collaboration with STScI, decided to produce the final calibration files for the science observations and the whole FOC dataset was re-calibrated accordingly \citep{Kamp2006}. From this data reprocessing, CADC noticed a sensible modification of the science data due to a new "best geometric reference." Since this re-calibration and the release of consistently processed data, no re-analysis of the observation has been published for the AGN dataset. In addition, most of the old reduced observations are dispersed throughout the literature, with different analyses and approaches, making it difficult to fully interpret the FOC AGN sample. This is why we decided to create a consistent, novel and open-access reduction pipeline of our own to produce high-level, science-ready, polarimetric products for the scientific community as well as polarimetric data reduction packages. Ultimately, we aim to explore, download, reduce, and present all the polarimetric images taken with the FOC in a standardized way. A large sample of radio-loud and radio-quiet AGNs is indeed necessary to investigate whether all the differences between pole-on and edge-on objects can be explained by an inclination effect \citep{Antonucci1993,Marin2016} or whether morphological differences of the circumnuclear region must also be taken into account \citep{Ramos2009,Alonso2011,Ichikawa2015,Lopez-Rodriguez2022}. Using a large AGN sample also allows us to study their thermal and non-thermal physical components \citep{Antonucci2012,Antonucci2015}, which, in turn, enables us to put physical constrains in the AGN components.

In this first paper of the series, we present a detailed description of the new reduction pipeline in Sect.~\ref{Pipeline}. We test our methodology against a well-known, previously published FOC polarimetric image of NGC~1068 in Sect.~\ref{NGC1068}. We then proceed with the reduction of a forgotten FOC observation of the Seyfert-2 galaxy IC~5063 (PKS~2048-572) in Sect.~\ref{IC5063}. We discuss our results Sect.~\ref{Discussion} and present our conclusions in Sect.~\ref{Conclusion}.

\section{Reduction pipeline}
\label{Pipeline}
\begin{figure*}
    \centering
    \includegraphics[width=\textwidth]{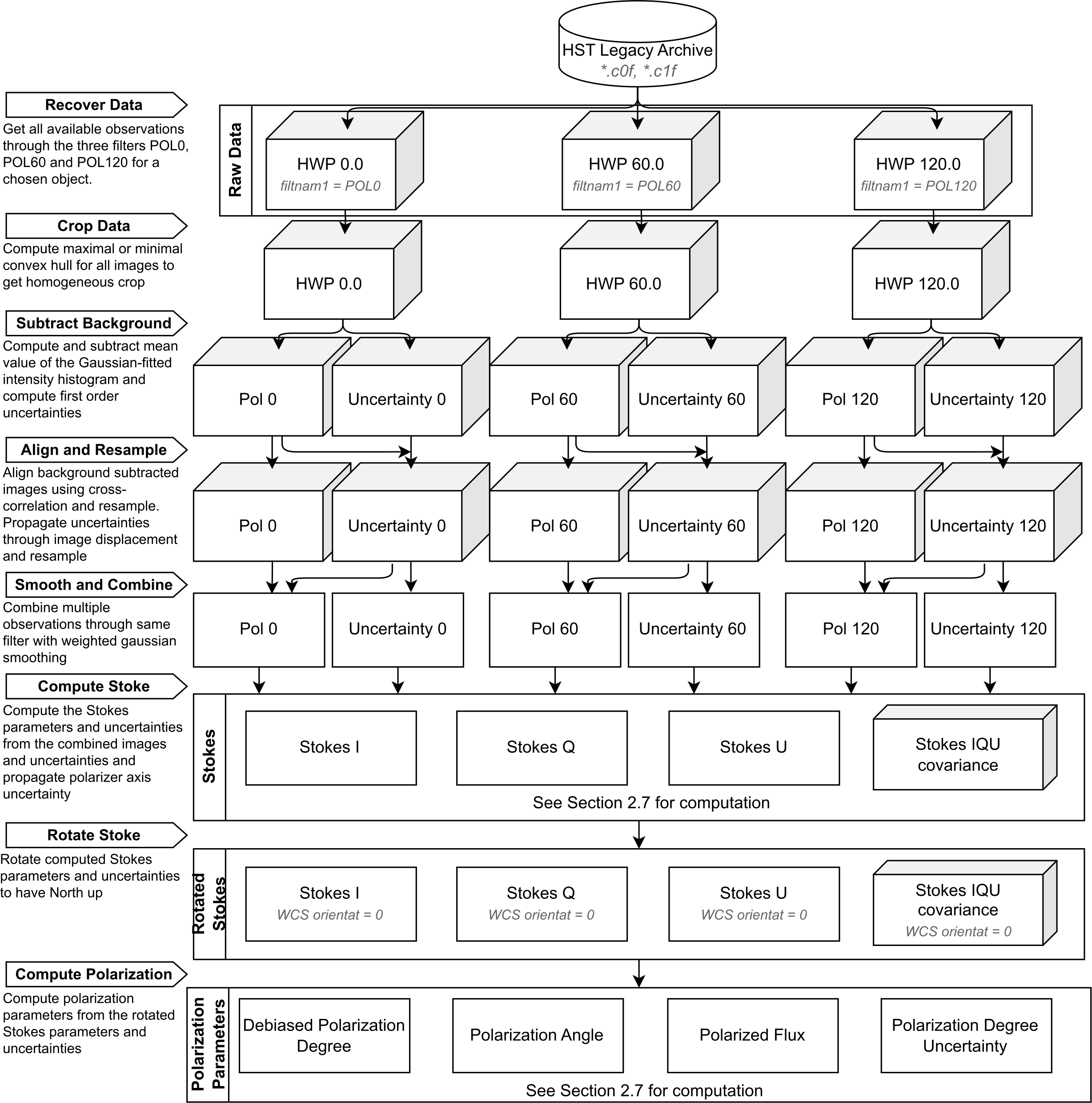}
    \caption{Diagram representing the pipeline's reduction operations from the raw data obtained from the HST Legacy Archives to the obtained polarization maps.}
    \label{fig:pipeline}
\end{figure*}

In this section, we present the general reduction pipeline methodology and our choice of various algorithms to extract as much information as possible from the raw data. We created our new reduction pipeline in \textsc{python} language, making use of this easy-to-read tool for optimized reduction methods. The pipeline is already available to grab on the author's \textsf{Git}\footnote{\href{https://git.unistra.fr/t.barnouin/FOC_Reduction}{git.unistra.fr/t.barnouin/FOC\_Reduction}}. We focused our work on the FOC instrument but it was written to be modular so that it is relatively easy to add other instruments to the pipeline. The overall data reduction steps is summarized in a diagram in Fig.~\ref{fig:pipeline}. The FOC instrument measures polarization state by performing three consecutive observations of the same target through three polarizer filters with complementary polarization axis angles, $\theta_1 = 0^{\circ}$, $\theta_2 = 60^{\circ}$, and $\theta_3 = 120^{\circ}$, usually referred to as POL0, POL60, and POL120 respectively. The reduction procedures of polarization observations with the FOC instrument call for at least three rounds of observations with the same instrument, but through three different filters with different properties \cite[see][Section 8.7]{DataHandbook1998}. The FOC instrument itself has some photometric uncertainties that also ought to be taken into account \cite[see][Section 8.3]{DataHandbook1998} as well as specific issues to the filter wheel, the uncertainty in the polarizer axis directions, PSF differences, and throughput issues \cite[see][Sections 4.4.3 and 11.2.6]{Nota1996} that induce relative uncertainties between observations. To better implement these uncertainties into our data reduction, we chose to go back to the most generic description of the flux through a polarizer, as described in the appendix of \citet{Kishimoto1999}. The  three polarized exposures obtained are then combined into the Stokes parameters to compute the polarization state, as described in Section \ref{Pipeline:Stokes}. For the remainder of this section, actual examples will be provided using the Feb 28, 1995 (5:33AM) FOC observation of NGC~1068 (ID: 5144), which we will explore in further detail in Sect.~\ref{NGC1068}.

\subsection{Data importing  and selection of the region of interest}
\label{Pipeline:Graham}
The data were imported from astrophysics standard FITS files, making use of information in headers to optimize the whole pipeline without requiring user inputs. The required FITS files were calibrated data products that can be retrieved from the MAST HST Legacy Archive\footnote{\href{https://archive.stsci.edu/missions-and-data/hst}{archive.stsci.edu/missions-and-data/hst}}. 

\begin{figure}
    \centering
    \includegraphics[width=\linewidth]{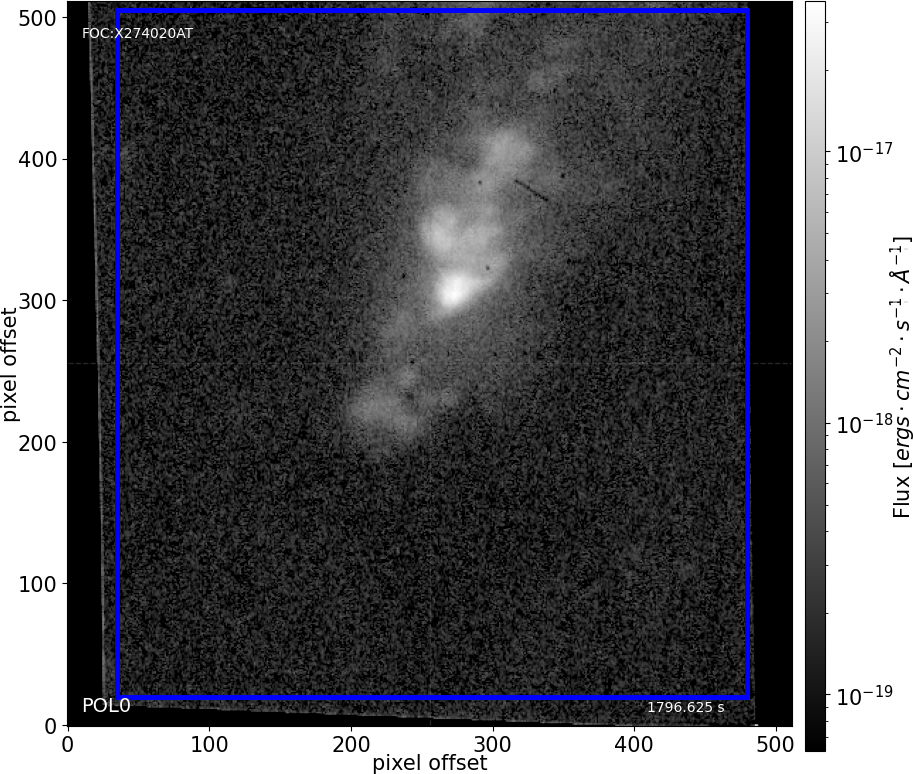}
    \caption{Image of the POL0 polarizer filter for NGC~1068. The total exposure time is 1,796 seconds. The blue rectangle delimits the cropped region of interest after Graham's scan, removing the borders containing non-pertinent data for the analysis.}
    \label{fig:NGC_1068_crop}
\end{figure}

A "query" utility that depends on \textsf{astroquery} Python package allows us to download FOC's \textsf{\_c0f} files from the terminal, given a target name (and possibly a proposal id). Otherwise, the user can feed its own FITS files to the pipeline, as long as their HEADER contain the identifying keywords defined in the HST/FOC Data Handbook Chapter 5.2 \citep{DataHandbook1998}. We made use of the Calibrated exposure FITS files, whose suffix is \textsf{\_c0f} in the MAST archives. We immediately translated each observation count as count rates, using the \textsc{EXPTIME} header keyword containing the exposure time of the data set. For a better handling of the data these count rates are conserved as such during the whole pipeline and only translated into physical units when displaying the relevant data through plots. This is done using the \textsc{PHOTFLAM} header keyword containing the inverse sensitivity conversion factor \citep{Nota1996}. In our case, count rates were transformed as fluxes in erg.cm$^{-2}$.s$^{-1}$.\AA$^{-1}$. For 2D images (e.g., FOC outputs), the observational data are processed through a Graham's scan algorithm \citep{Graham1972} for a better selection of the region of interest (ROI). This algorithm finds the convex hull of a set of $n$ points in the plane with a complexity $O\left(n \log n \right)$, cropping out non-exploitable values from the data matrix (infinite numbers, zeros, ...). An automatized function finds the optimal rectangle-shaped image that contains only valuable data from the observation and removes the unusable empty borders, artifacts from the finite size of the detector and non-physical calibration procedures. The implemented version of this algorithm takes the full set of observations and concurrently crop out the undesired edges on every observation. To do so, it takes as the parameters the pixel step to create the image shell (allows us to reduce the number of points to be considered to run the algorithm), the value to be discarded, and the choice of whether the final crop should be the intersection or the union of all individual crops. The obtained shell is then intersected with the ones obtained for each observations (and each half wave plate) to get an uniform cropping across the whole dataset. An example of such preliminary data selection can be seen in Fig.~\ref{fig:NGC_1068_crop}. On average, this procedure remove 15 -- 18\% of the original raw image of  512 $\times$
512 pixels.

\subsection{Deconvolution.}
\label{Pipeline:Deconvolution}
Before the installation of COSTAR, the FOC point spread function (PSF) suffered from severe spherical aberration, which meant that a circular aperture of 0.1" radius contained only 15 -- 18\% of the light from a star instead of the expected 70\%. COSTAR has restored much of the OTA capabilities, in the sense that the COSTAR-corrected PSF contains more than 75\% of the light within a radius of 0.1" at visible wavelengths. The FOC PSF typically measures 1.6 -- 1.8 pixels in the visible ($\sim$ 0.08" at full width at half maximum, FWHM), as in \citet{Nota1996}. To recover the underlying fine structures that has been blurred by the photo-diodes of the detector, our pipeline implements several deconvolution algorithms that can be used to treat the raw images before any reduction. 

A linear regularized deconvolution method was implemented using a standard Wiener filter \citep{Wiener1949}. It is a low-pass filter with some neighboring pixel regularization constraint. Its simplicity is optimal for stationary and Gaussian noise, but for spatially localized features such as singularities or edges, it comes with drawbacks; namely, it creates oscillations along the sharp contours and degrades the resolution. 

Several iterative regularized methods were also implemented, which enforce a set of additional constraints of positivity, support, and band limitation on a given object, $O$. Because this process investigates the maximum likelihood in the case of Poisson-noise induced by a PSF $P$, it has no closed-form solution and requires an iterative approach. The algorithm evaluates the most probable pixel in which a photon should be detected given the raw image, $I$. The Van-Cittert method finds underlying structures in residual solutions at each iteration and put them in the restored image: $O^{n+1} = O^{n} + \alpha (I - (P * O^n)),$ where $\alpha$ is a convergence parameter generally taken to be equal to $1$ \citep{vanCittert1931}. The one-step gradient method replaces this convergence parameter with a convolution of the residual with the inverted PSF : $O^{n+1} = O^{n} + P^* * (I - (P * O^n))$. The Richardson-Lucy method multiplies the previously computed deconvolved image with a weighted image made up of a convolution of the data with the known PSF of the detector : $O^{n+1} = O^{n} (\frac{I}{P * O^n} * P^*)$ \citep{Richardson1972}. Finally, the conjugate gradient iterative method solves the inverse problem with PSF convolution and regularization constraint in an optimized way: the search direction for the solution $O^n$ is orthogonal to the direction of the gradient of the residual function $R^n(x,y) = (I - P * O^n)(x,y)$. 

In our pipeline, we let the user decides if a deconvolution should be applied and how many times any iterative algorithm should be run. This is critical for pre-COSTAR observation but less important in the case of post-COSTAR data, as explained previously. In our example of NGC~1068, no deconvolution was performed on the data.

\subsection{Error computation and propagation.}
\label{Pipeline:Error}
\begin{figure}
    \centering
    \includegraphics[width=\linewidth]{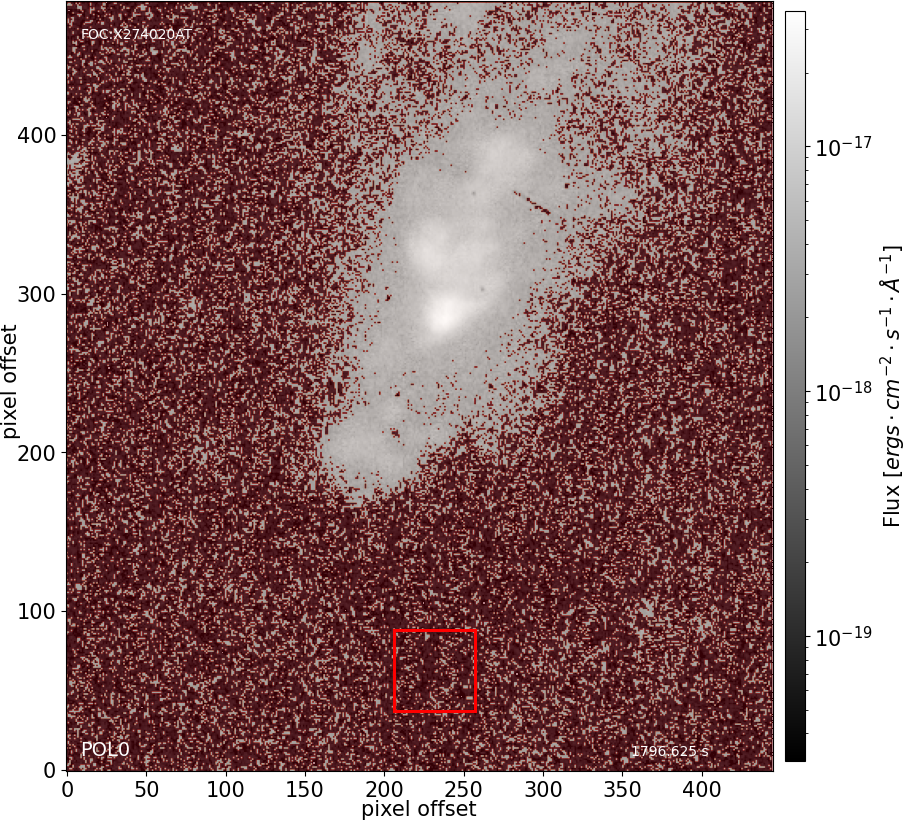}
    \caption{Image of NGC~1068 for the first observation. The red rectangle delimits the region considered for background noise. The dark red pixels are considered to be below the background intensity value.}
    \label{fig:NGC_1068_error_location}
\end{figure}

\begin{figure}
    \centering
    \includegraphics[width=\linewidth]{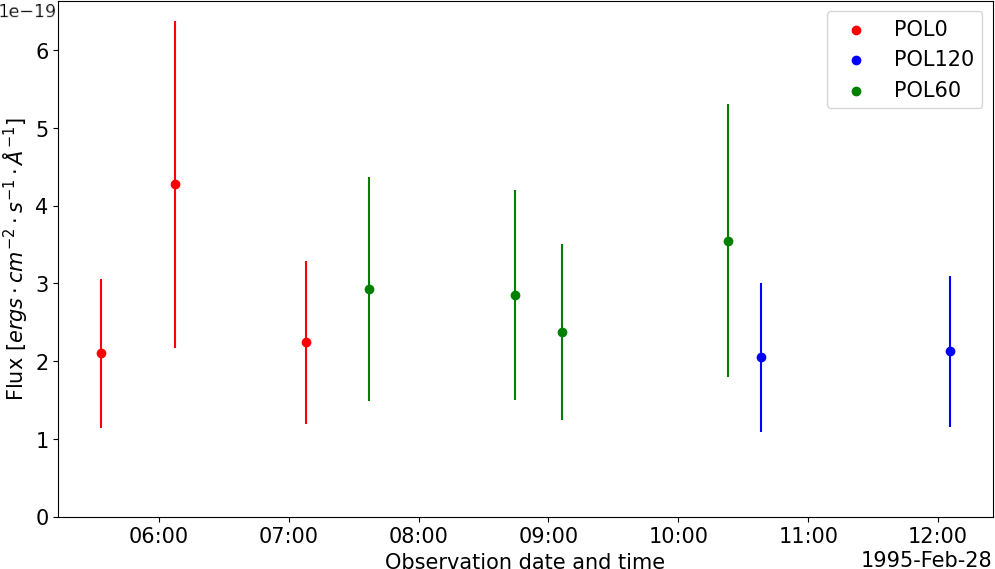}
    \caption{Background flux and error for each NGC~1068 dataset as a function of the observation time. The different colors represent the various polarizer filters (polarization axis angle of 0$^{\circ}$, 60$^{\circ}$ and 120$^{\circ}$ for the FOC instrument).}
    \label{fig:NGC_1068_error}
\end{figure}

The background noise is estimated on the calibrated data, before being processed through data alignment and resampling. A very basic first method searches for a common region in all observations and of user defined pixel size (basically, $\sim$10\% of the image size) with the least integrated flux. We assume this sub-image to be background dominated and we estimate the background by taking the root mean square of the selected sub-image (see Fig.~\ref{fig:NGC_1068_error_location}). The user can check for the evolution of the background flux during each observation (see Fig.~\ref{fig:NGC_1068_error}) and verify that there is no transient source involved. 

\begin{figure}
    \centering
    \includegraphics[width=\linewidth]{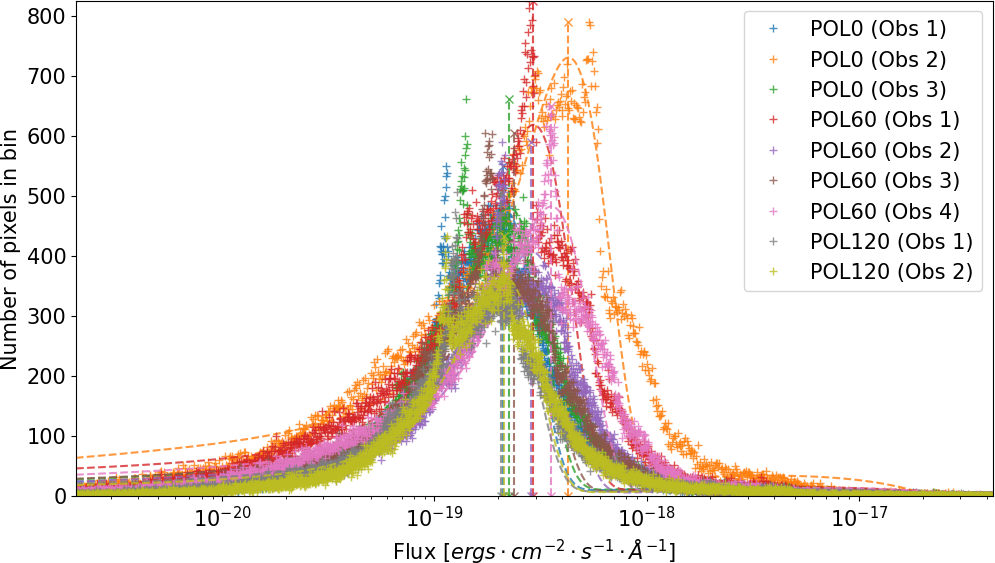}
    \caption{Intensity histograms on which the background (vertical dashed line) is estimated for each observation.}
    \label{fig:NGC_1068_error_histograms}
\end{figure}

Another more robust method takes into account the intensity histogram of each image and assume that the background is the most represented intensity bin (see Fig.~\ref{fig:NGC_1068_error_histograms}). The binning is done with a logarithmic range, in such a way that lower intensities get more precise binning than high intensities, and the number of bins is given by the Freedman-Diaconis rule for a sample, $x,$ of size, $n$: 
\begin{equation}
    N_{bins} = \frac{\max(x) - \min(x)}{2 \cdot \frac{IQR(x)}{\sqrt[3]{n}}}, \hfill \text{\footnotesize with $IQR$ the interquartile range.}
\end{equation}
If it is required by other flux statistics during the observations, the user can also choose for the number of bins to be computed by the following rules: square-root ($N_{bins}=\sqrt(n)$), Sturges ($N_{bins}=\log_2(n)+1$), Rice ($N_{bins}=2\sqrt[3]{n}$), and Scott ($N_{bins}=\frac{\max(x)-\min(x)}{3.5 \frac{STD(x)}{\sqrt[3]{n}}}$).

This second method has little dependence on transient sources as it looks for an intensity plateau rather than an image location. It can also better estimate observation-dependent levels, as can be seen in the deviation of estimated background values in Table~\ref{tab:NGC_1068_histogram_bkg}. These differences in intensity levels can come from different parameters of observation or calibration. At this point, the background value is subtracted to the whole image. From there are computed and quadratically summed the uncertainties in the "correction factors" as a percentage of the observed flux in each pixel. Following \citet{Kishimoto1999}, the wavelength dependence of the transmittance of each polarizer filter $\sigma_{wav}$ is taken to be 1\%, the differences in PSFs through each polarizer filter, $\sigma_{psf}$, is taken to be 3\% and the heavily smoothed flat-fielding uncertainty, $\sigma_{flat}$, is taken to be 3\%.

\begin{table}
    \centering\small
    \resizebox{\linewidth}{!}{%
        \begin{tabular}{c|c|c|c}
            Filter name & Observation date \& time & $N_{bins}$ & Estimated background ($10^{-4} s^{-1}$)  \\ 
            \hline
            POL0        & 1995-02-28 05:33:12   & 6348       & 6.43                         \\
            POL0        & 1995-02-28 06:07:33   & 2141       & 25.7                         \\
            POL0        & 1995-02-28 07:07:54   & 5747       & 7.78                         \\
            POL60       & 1995-02-28 07:37:16   & 4319       & 13.1                         \\
            POL60       & 1995-02-28 08:44:26   & 6091       & 10.3                         \\
            POL60       & 1995-02-28 09:06:15   & 6205       & 8.57                         \\
            POL60       & 1995-02-28 10:22:57   & 4727       & 16.3                         \\
            POL120      & 1995-02-28 10:37:59   & 8324       & 7.06                         \\
            POL120      & 1995-02-28 12:05:37   & 8696       & 6.49
        \end{tabular}%
    }
    \caption{Histogram binning and resulting estimation of the background (here in count rates) for each observation of NGC~1068.}\vspace{-15pt}
    \label{tab:NGC_1068_histogram_bkg}
\end{table}

\subsection{Data alignment.}
\label{Pipeline:Alignment}
The polarized data come from multiple observations with different polarizer filter. To extract the polarization information in the Stokes convention, we must sum the datasets, but it is only possible if the data have been previously aligned. To do so, our method implements a 2D image alignment to sub-pixel precision using a factor of 10 oversampling, allowing us to align our images to a precision up to 0.1 pixel precision, corresponding to an alignment precision between observations of 0.0014 arcsec. This is done through cross-correlation of the phase-space of the misaligned images, as described in \citet{GuizarSicairos2008}. Each image is then linearly shifted accordingly, using the relation in Eq.~\ref{eq:shift} to compute the value in each pixel. Once aligned using the imaged large scale structures, the uncertainty coming from the different observations shifts is computed from the displacement with respect to the reference dataset. This uncertainty is computed for each pixel in the resulting image, as half of the difference of the values in this pixel before and after shifting the data (see Eq.~\ref{eq:shifterror}). This uncertainty is quadratically summed to the global uncertainty inside the pixel:

\begin{gather}
    \begin{aligned}
        I_{x,y}^{shifted}(\Delta x,\Delta y) = &uv \cdot I_{x+\lfloor{\Delta x}\rfloor,y+\lfloor{\Delta y}\rfloor}\label{eq:shift}\\
        &+ u(\mathrel{{1}{-}{v}}) \cdot I_{x+\lfloor{\Delta x}\rfloor,y+\lceil{\Delta y}\rceil}\\
        &+ (\mathrel{{1}{-}{u}})v \cdot I_{x+\lceil{\Delta x}\rceil,y+\lfloor{\Delta y}\rfloor}\\
        &+ (\mathrel{{1}{-}{u}})(\mathrel{{1}{-}{v}}) \cdot I_{x+\lceil{\Delta x}\rceil,y+\lceil{\Delta y}\rceil}
    \end{aligned},\\
    \sigma_{x,y}^{shifted}(\Delta x,\Delta y) = \left|\frac{{I_{x,y} - I_{x,y}^{shifted}(\Delta x,\Delta y)}}{2}\right|,\label{eq:shifterror}\\
    \nonumber \text{with } \left\{\begin{aligned}
        &\Delta x, \Delta y \text{ are the shifts along the x and y axis},\\
        &\lfloor \Delta x \rfloor, \lceil{\Delta x}\rceil \text{ the floor and ceiling integers of } \Delta x,\\
        &u = \Delta x - \lfloor \Delta x \rfloor, v = \Delta y - \lfloor \Delta y \rfloor.
    \end{aligned}\right.
\end{gather}

\subsection{Data binning}
\label{Pipeline:Resampling}
We propose several methods to re-sample the data. Assuming the target pixel size is larger than the original pixel size, the user can resample the data in \textit{arcsec} or \textit{pixel} units and can choose to do it by averaging or summing the resampled data. This is done by re-binning the data matrix to a smaller shape using matrix products with rows and columns compressors. This resampling, while reducing the spatial resolution, allows us to get better statistics in the resized pixel that now sums or averages the events from each sub-pixels. This is mandatory to study polarization as polarized fluxes require high statistics to become meaningful. In the extreme case scenario, the user can integrate the whole image down to one pixel to simulate what a polarimetric instrument without imaging capabilities would have observed. Uncertainties computed from previous alignment and background subtraction procedures are propagated through re-sampling as the quadratic sum of the errors of the bin. This uncertainty is then quadratically summed to the root mean square (RMS) of the flux of the sub-pixels of the bin, accounting for some baseline noise:
\begin{gather}
    \sigma_{X,Y}^{re-sampling} = \sqrt{{RMS_{X,Y}^f}^2 + {\sigma_{X,Y}^{propag}}^2},\\
    \nonumber \text{with } \left\{
    \begin{aligned}
        RMS_{X,Y}^f &= \frac{\sqrt{\sum_{(x,y) \in (X,Y)} f_{x,y}^2}}{\sum_{(x,y) \in (X,Y)} 1},\\
        \sigma_{X,Y}^{propag} &= \sqrt{\frac{\sum_{(x,y) \in (X,Y)} \sigma_{x,y}^2 \cdot f_{x,y}}{\sum_{(x,y) \in (X,Y)} f_{x,y}}},
    \end{aligned}\right.
\end{gather}
where each pixel of the re-binned image at coordinates, $X,Y,$ correspond to a subset of pixels $(x,y) \in (X,Y)$ in the original image of flux, $f_{x,y}$, and associated error, $\sigma_{x,y}$.

\subsection{Data smoothing}
\label{Pipeline:Smoothing}
Several options are available for smoothing the data. The idea behind data smoothing is to reduce noise from a data set to allow important patterns to more clearly stand out. A user-defined function can be convolved to the prepared data, before summing observations that were obtained through the same polarizer filter. The same convolution procedure can also be done after data combination. This convolution can be applied to a weighted dataset whose weights, $w_{i,j}$, are the inverse square of the error for each pixel $x,y$:
\begin{equation}
    S_{xy} = (s * g)_{xy} = \sum_{i,j}^{N_{pixels}} s_{ij} \cdot w_{ij} \cdot g(\mathrel{{x}{-}{i}},\mathrel{{y}{-}{j}}),
\end{equation}
where $g$ is some user-defined kernel to which the data should be convoluted. The error of the smoothed pixel is computed by convoluting the square of the errors to the convolution kernel squared:
\begin{equation}
    \sigma_{xy}^{smoothing} = \sqrt{(\sigma^2 * g^2)_{xy}} = \sqrt{\sum_{i,j}^{N_{pixels}} \sigma_{ij}^2 \cdot w_{ij}^2 \cdot {g(\mathrel{{x}{-}{i}},\mathrel{{y}{-}{j}})}^2}
.\end{equation}

Another finer data smoothing combines and smooth the data from multiple observation sets at the same time using a Gaussian kernel with user-defined FWHM. Given $N$ observations through a given polarizer filter, the obtained combined and smoothed pixel at coordinates $(x,y)$ is given by:
\begin{equation}
    S_{xy} = \frac{\sum_k^{N}\sum_{i,j} s^k_{ij} \cdot w^k_{ij} \cdot g(\mathrel{{x}{-}{i}},\mathrel{{y}{-}{j}})}{\sum_k^{N}\sum_{i,j} w^k_{ij} \cdot g(\mathrel{{x}{-}{i}},\mathrel{{y}{-}{j}})}
,\end{equation}
where $s^k_{ij}$ is the signal of the pixel at $(i,j)$ for observation $k$, $g(\mathrel{{x}{-}{i}},\mathrel{{y}{-}{j}}) = e^{-\frac{(\mathrel{{x}{-}{i}})^2+(\mathrel{{y}{-}{j}})^2}{2 \sigma^2}}$ is a Gaussian kernel with $\sigma = \text{FWHM} / (2\sqrt{2 \ln{2}}),$ and $w^k_{ij} = 1/{e^k_{ij}}^2$ is the weight given by the inverse-squared error of this same pixel. The error on the combined pixel is obtained by taking the weighted root mean square of the errors:
\begin{equation}
    \sigma_{xy}^{smoothing} = \frac{\sqrt{\sum_k^{N}\sum_{i,j} {e^k_{ij}}^2 \cdot {w^k_{ij}}^2 \cdot {g(\mathrel{{x}{-}{i}},\mathrel{{y}{-}{j}})}^2}}{\sum_k^{N}\sum_{i,j} w^k_{ij} \cdot g(\mathrel{{x}{-}{i}},\mathrel{{y}{-}{j}})}
.\end{equation}

\subsection{Stokes parameters and polarization components computation.}
\label{Pipeline:Stokes}
Here, we  describe how we computed the Stokes parameters and the uncertainties arising from the polarization measurement. We refer to \citet{Kishimoto1999} and \citet[p38, p100, p171]{ClarkeGrainger1972} for more details.
The FOC instrument is not equipped with a circular polarization analyzer that would allow us to characterize the ellipticity of the observed polarization. In the following, we assume a linear polarization, with a Stokes parameter of $V = 0$.

We computed the remaining $I$, $Q$, $U$ Stokes parameters from the addition and subtraction of the theoretical flux through three polarizer filters with complementary polarization axis angles ($\theta_1 = 0^{\circ}$, $\theta_2 = 60^{\circ}$, $\theta_3 = 120^{\circ}$ for the FOC). We defined the Stokes vector as $\mathbf{S} = (I, Q, U)$. We call $F_i$ the theoretical polarized flux that is not attenuated by the polarizer filter with axis angle $\theta_i$. We define the theoretical polarized flux vector by $\mathbf{F} = \left(\frac{2 f_{\theta_1}}{t_1}, \frac{2 f_{\theta_2}}{t_2}, \frac{2 f_{\theta_3}}{t_3}\right),$ where $t_i$ is the transmittance of the polarizer filter with axis oriented to angle $\theta_i$. The general formula is the following: $\mathbf{S} = A \cdot \mathbf{F}$ and the transformation matrix, $A,$ is given by :
\begin{gather}
    A = \frac{1}{N} 
    \left[\begin{smallmatrix}
        k_2 k_3 \sin{(-2\theta_2 + 2\theta_3)} & k_3 k_1 \sin{(-2\theta_3 + 2\theta_1)} & k_1 k_2 \sin{(-2\theta_1 + 2\theta_2)}\\
        -k_2 \sin{2\theta_2}+k_3\sin{2\theta_3} & -k_3 \sin{2\theta_3}+k_1\sin{2\theta_1} & -k_1 \sin{2\theta_1}+k_2\sin{2\theta_2}\\ 
        k_2 \cos{2\theta_2}-k_3\cos{2\theta_3} & k_3 \cos{2\theta_3}-k_1\cos{2\theta_1} & k_1 \cos{2\theta_1}-k_2\cos{2\theta_2}
    \end{smallmatrix}\right]\label{eq:StokesMatrix},\\
    \begin{aligned}
        \nonumber\text{ where \;} N = &k_2 k_3 \sin{(-2\theta_2 + 2\theta_3)} + k_3 k_1 \sin{(-2\theta_3 + 2\theta_1)}\\
            &+ k_1 k_2 \sin{(-2\theta_1 + 2\theta_2).}
    \end{aligned}
\end{gather}

We then define $A'$ such that $\mathbf{S} = A' \cdot \mathbf{f}$ where $\mathbf{f} = (f_{\theta_1},f_{\theta_2},f_{\theta_3})$ is the observed polarization flux vector.

The error is propagated through the transformation of the variance-covariance matrix of the polarization flux, $\mathbf{f}$ ($V^{\mathbf{f}}$), to that of the Stokes parameters $\mathbf{S}$  ($V^{\mathbf{S}}$), where we assumed no correlation between the flux obtained through the different polarization filters: 
\begin{gather}
    V^{\mathbf{S}} = A' V^{\mathbf{f}} A'^{T},\\
    \nonumber \text{\; with \;}
    V^{\mathbf{S}} = \left[
    \begin{smallmatrix}
        \sigma_I^2 & \sigma_{IQ} & \sigma_{IU}\\
        \sigma_{IQ} & \sigma_Q^2 & \sigma_{QU}\\
        \sigma_{IU} & \sigma_{QU} & \sigma_U^2
    \end{smallmatrix}\right] \text{ , } 
    V^{\mathbf{f}} = \left[
    \begin{smallmatrix}
        \sigma_{f_{\theta_1}}^2 & 0 & 0\\
        0 & \sigma_{f_{\theta_2}}^2 & 0\\
        0 & 0 & \sigma_{f_{\theta_3}}^2
    \end{smallmatrix}\right].
\end{gather}

The statistical uncertainty is computed after the combination of the observed polarized flux for each Stokes parameter. The uncertainty on the polarized flux $\sigma_{f_{\theta_j}}$ is calculated assuming Poisson noise in the counts and $\forall k \neq j, \, \sigma_{f_{\theta_j}f_{\theta_k}} = 0,$ as they arise from different observations:\begin{alignat}{2}
    {\sigma^{stat}_{S_i}}^2 &= \sum_{j=1}^3 \left|\frac{\partial S_i}{\partial f_{\theta_j}}\right|^2 \cdot \sigma^2_{f_{\theta_j}} \;&&\text{ for } S_i \in \left[I, Q, U\right],\\
    \sigma_{f_{\theta_j}} &= \sqrt{\frac{r_j}{t_j}} \;&&\text{ for } f_{\theta_j} \in \left[f_{\theta_1},f_{\theta_2},f_{\theta_3}\right],
\end{alignat}
with $\forall j$, $r_j$ represent the rate and $t_j$ is the exposure time for the polarized flux, $f_{\theta_j}$. We compute the partial derivative of $S_i$ with respect to $f_{\theta_j}$, knowing that $\forall j, \frac{\partial A'}{\partial f_{\theta_j}} = 0$:
\begin{equation}
    \frac{\partial S_i}{\partial f_{\theta_j}} = \sum_{k=1}^3 {A'}_{ik} \frac{\partial f_{\theta_k}}{\partial f_{\theta_j}} = A'_{ij}
.\end{equation}

The polarizer filters axis angle is known to an uncertainty of $3^{\circ}$ \citep{Nota1996} and this error comes into account when computing the Stokes parameters as they explicitly depend on $\theta_{1,2,3}$ through the transformation matrix (Eq. \ref{eq:StokesMatrix}). Assuming $\sigma_{\theta} = 3^{\circ}$, we compute the uncertainties from the polarizer filters orientation as follows:
\begin{equation}
    {\sigma^{axis}_{S_i}}^2 = \sum_{j=1}^3 \left|\frac{\partial S_i}{\partial \theta_j}\right|^2 \cdot \sigma^2_{\theta_j} \;\text{ for } S_i \in \left[I, Q, U\right]
,\end{equation}
where we compute the partial derivative of $S_i$ with respect to $\theta_j$ assuming $\forall k \neq j, \frac{\partial f_{\theta_k}}{\partial \theta_j} = 0$:
\begin{equation}
    \frac{\partial S_i}{\partial \theta_j} = \frac{1}{N} \left[ \sum_{k=1}^3 \frac{\partial a'_{ik}}{\partial \theta_j} f_{\theta_j} - S_i \frac{\partial N}{\partial \theta_j}\right] \text{\; with } a' = N \cdot A'
.\end{equation}
These uncertainties are quadratically summed to the previously computed propagated errors.

We then rotated Stokes parameters to have north directed up. From the header keyword \textsc{ORIENTAT} providing the angle between north and the image's y axis in the northeast direction, we get $\alpha$ the rotation angle. We transform the Stokes parameters and covariance matrix in the following way :

\begin{gather}
    \mathbf{S}_r = R_I\left(-2\alpha\right) \cdot \mathbf{S},\\
    V^{\mathbf{S}}_r = R_I\left(-2\alpha\right) \cdot V^{\mathbf{S}} \cdot {R_I\left(-2\alpha\right)}^{T},\\
    \nonumber \text{where } R_I\left(-2\alpha\right) = \left[
    \begin{matrix}
        1 & 0 & 0\\
        0 & \cos{(2\alpha)} & \sin{(2\alpha)}\\
        0 & -\sin{(2\alpha)} & \cos{(2\alpha)}
    \end{matrix}\right]. 
\end{gather}

The polarization degree and angle are determined from the Stokes parameters by the following well-known equations:
\begin{align}
    P &= \frac{\sqrt{Q^2 + U^2}}{I},\\
    \theta_P &= \frac{1}{2} \arctan \left( \frac{U}{Q} \right).
\end{align}
And the associated errors are propagated as follows:
\begin{alignat}{2}
    \sigma_P &= \frac{1}{I} && \left(\frac{Q^2 \sigma_Q^2 + U^2 \sigma_U^2 +2QU \sigma_{QU}}{Q^2 + U^2} + \frac{Q^2+U^2}{I^2} \sigma_I^2 \right.,\\
    \nonumber & && \; \left. - \frac{2Q}{I} \sigma_{IQ} - \frac{2U}{I} \sigma_{IU} \right)^\frac{1}{2}\\
    \sigma_{\theta_P} &= &&\frac{1}{2 \left(Q^2+U^2\right)} \left(U^2 \sigma_Q^2 + Q^2 \sigma_U^2 -2 QU \sigma_{QU}\right)^\frac{1}{2}.
\end{alignat}

Due to the presence of noise, the normalized Stokes parameters are the only estimates of the true normalized Stokes parameters. To correct this bias, in the following, we refer to the polarization degree as its improved estimator: the debiased polarization degree of $P_{debiased} = \sqrt{P^2 -\sigma_P^2}$ \citep{Simmons1985}.

\section{Benchmarking our pipeline against NGC~1068}
\label{NGC1068}
In order to test our pipeline, we decided to re-analyze the FOC data of NGC~1068. This is the most archetypal type-2 (edge-on) radio-quiet AGN and thus the best target for benchmarking. It possesses the largest database of radio-to-UV polarization measurements \citep{Marin2018b} and was even part of the original catalog of Carl Seyfert \citep{Seyfert1943}. Its proximity to Earth ($z \approx$ 0.00379, which corresponds to a Hubble distance of $\sim$ 13.48~Mpc in the standard $\Lambda$CDM model) allows us to resolve the first hundreds of parsecs thanks to the spatial resolution of the FOC (at $z$ = 0.00379, 1" equals 81.5~pc). NGC~1068 was observed by the FOC on Sep 30, 1993 (10:32AM) and on Feb 28, 1995 (5:33AM), with the respective program IDs 3504 and 5144. However, the first observation (1993) was badly saturated along the AGN core direction due to an improper estimation of the expected UV flux \citep{Capetti1995a}. We thus focus our work on the second observation (1995). The dataset was obtained through the F253M UV filter centered around 2530~\AA, together with the polarizing filters POL0, POL60, POL120. The optical relay f/96 was selected to obtain a FoV of 7" $\times$ 7" and a pixel size of 0.014" $\times$ 0.014". It results in a 512 $\times$ 512 pixelated image of the source and its environment. Each polarizing filter acquired $\sim$ 3\,500 seconds worth of observation for a total exposure time of 10\,581 seconds. This observation was first published by \citet{Capetti1995b} and further analyzed by \citet{Kishimoto1999}.

\begin{figure*}
    \centering
    \includegraphics[width=\textwidth]{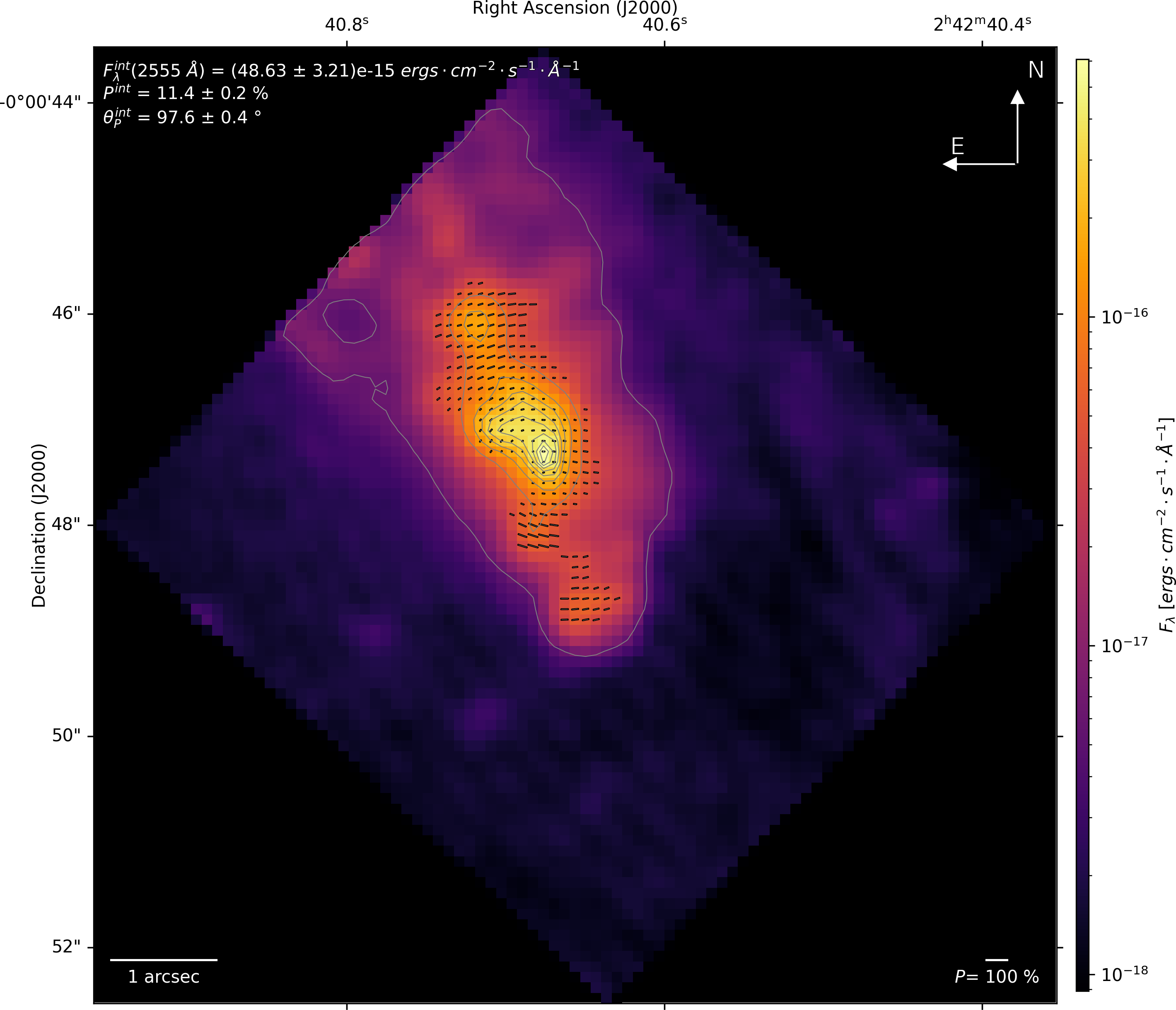}
    \caption{ Total flux F$_\lambda$ (erg.s$^{-1}$.cm$^{-2}$.\AA$^{-1}$, color-coded) of NGC~1068, with the polarization information superimposed to the image using white vectors. The linear polarization degree is proportional to the vector length while the polarization position angle is indicated by the orientation of the vector (a vertical vector indicating a polarization angle of 0$^\circ$). North is up, east is left. We show the full FoV so that no potential information is lost. A spatial bin corresponds to 0.1".  The contours are displayed from 1\% to 99\% every 10\% of the maximum value of $ 4.863\cdot 10^{-14}$ erg.s$^{-1}$.cm$^{-2}$.\AA$^{-1}$. The smoothness of the contours is tightly linked to the amount of smoothing done in the reduction process.}
    \label{fig:NGC_1068}
\end{figure*}

In Fig.~\ref{fig:NGC_1068}, we show the total intensity map with the polarization information superimposed to it. We rebinned the data in order to get a pixel size of 0.1", similarly to \citet{Capetti1995b} for a direct comparison. The pixels were smoothed by a Gaussian kernel of standard deviation 0.2". Only the bins with a S/N higher than 30 in total flux and higher than 3 in polarization have their polarization vectors shown. We display in the top-left corner of the figure the total flux, polarization degree, and polarization angle as integrated over the whole FoV. At 2555\AA\ pivot wavelength we observe an integrated flux of $(48.63 \pm 3.21) \cdot 10^{-15}$ erg.s$^{-1}$.cm$^{-2}$.\AA$^{-1}$ with a polarization of $11.4 \pm 0.2$ \% at position angle $97.6 \pm 0.4 ^{\circ}$. The total flux image is dominated by a compact region of about 3 $\times$ 3 pixels (0.3" $\times$ 0.3") that is situated at the base of the polar outflows that extend in the northern and southern directions, although the northern part is much more visible in total flux. The southern part suffers from slightly higher reddening from the dust of the host galactic plane \citep{Kishimoto1999}. The polarization vectors allow to highlight the double-conical morphology of the winds, with a much higher contrast than in pure intensity. The polarization pattern seen in the winds is centro-symmetric, pinpointing the source of emission even if it is hidden by an optically thick dusty region. This circumnuclear region, often called the "torus", is not coincident with the brightest spot of total flux but is immediately below it ($\sim$ 0.3" towards the south), where the total intensity decreases due to heavy absorption by the dust and gas mixture \citep{Kishimoto1999}. Focusing on the regions with the highest S/N, the structure of the bi-polar winds is revealed thanks to polarimetry. The polarization degree in the brightest region is of the order of 15 -- 20\% and significantly increases along the winds, up to 35 -- 40\%.

\begin{figure}
    \centering
    \includegraphics[width=\linewidth]{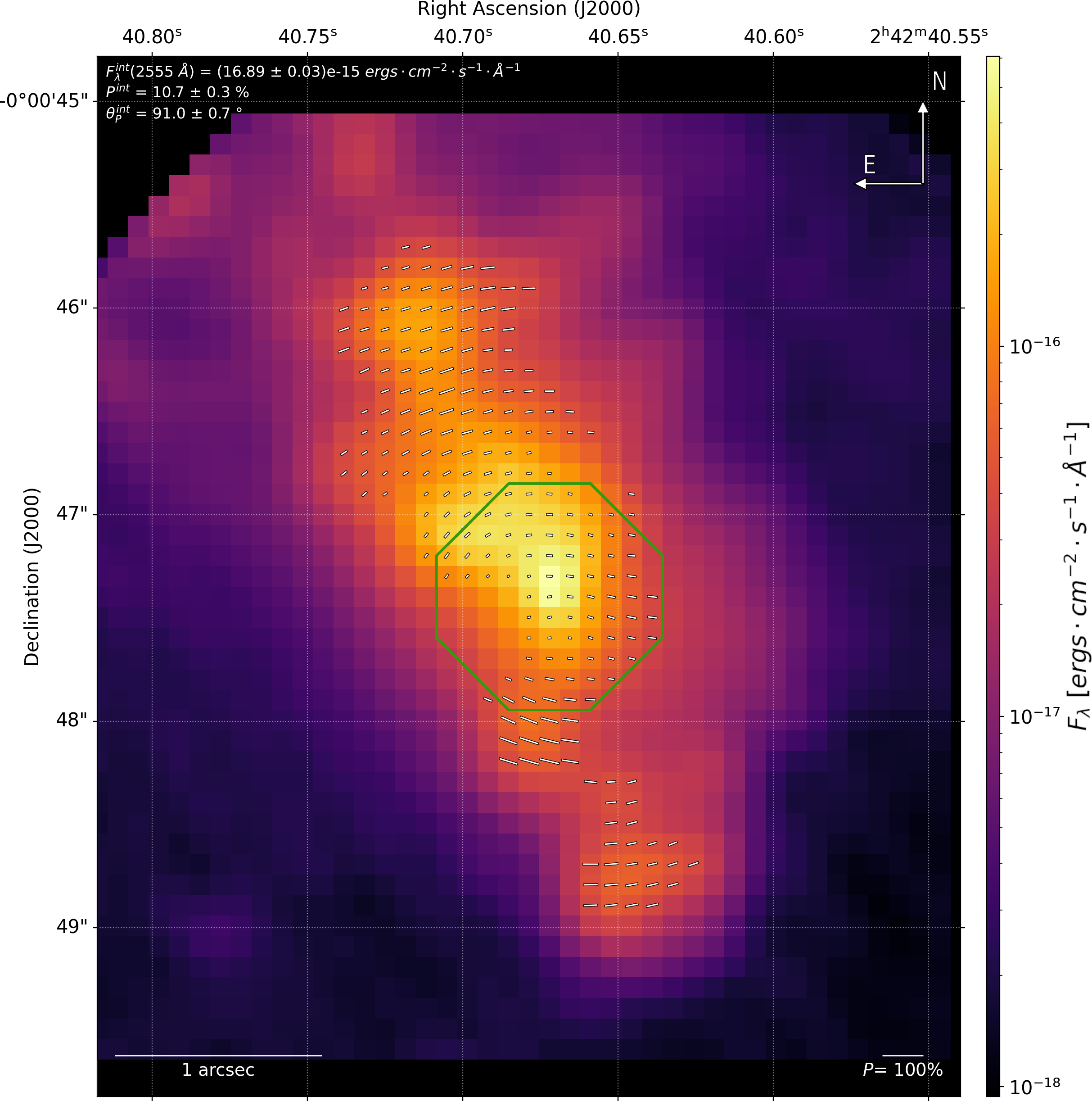}
    \caption{Comparison of the integrated polarization obtained by this pipeline through a simulated aperture of 1" diameter (green encircled region) to the one obtained by \citet{Antonucci1994} with the FOS spectropolarimeter. The FOC data was binned to a pixel of 0.1" and smoothed with a Gaussian of a FWHM of 0.2". The polarization vectors are shown for $\left[\text{S/N}\right]_P \geq 3$ and $\left[\text{S/N}\right]_I \geq 30$.}
    \label{fig:comparison_A}
\end{figure}

A direct comparison with the results of \citet{Capetti1995b} validates ours results. The centro-symmetric pattern is well reproduced and the differences in morphology observed between the northern and southern outflows coincides between Fig.~2 in \citet{Capetti1995b} and our Fig.~\ref{fig:NGC_1068} with a S/N cut of 30 in total flux. We find similar polarization degrees in various positions in the winds, with the exception of the location of the highest patch of polarization degree. In \citet{Capetti1995b}, the authors detect a 65\% polarization level immediately west of the brightest spot of total flux, while such this region only displays 20\% linear polarization in our results. This is very likely due to the different steps between the two reduction pipelines. While in \citet{Capetti1995b} the errors in the Stokes parameters were computed assuming Poisson statistics, we propagated the errors from the calibrated data from the archive. In addition, the smoothing of the image can play a crucial role in blurring or increasing the polarization of a given pixel. No indication on the smoothing process were given in \citet{Capetti1995b}, but we tried using different Gaussian kernels without succeeding to find the same patch of high polarization. On the other hand, we note that \citet{Kishimoto1999} re-analyzed the same data and neither found this 65\% polarization spot. It indicates that there might have been a small misalignment effect or a numerical artifact in the reduction method of \citet{Capetti1995b}. There are, in fact, no evident reasons why such a high polarization degree should exist outside the polar winds half-opening angle, directly west of the torus location, where the central irradiation should be absorbed by the circumnuclear dust wall. We also note that we present a much wider view of NGC~1068, while \citet{Capetti1995b} cropped their results to a FoV of 3.3" $\times$ 2.9". Integrating the total flux over the 7" $\times$ 7" image with a binning of 0.10" and a Gaussian combination smoothing of a FWHM of 0.20" gives us about 4.86 $\pm$ 0.32 $\times$ 10$^{-14}$ erg.s$^{-1}$.cm$^{-2}$.\AA$^{-1}$, which agrees with the 4.79 $\pm$ 0.20 $\times$ 10$^{-14}$ erg.s$^{-1}$.cm$^{-2}$.\AA$^{-1}$ flux recorded by the International Ultraviolet Explorer (IUE) for a 10" $\times$ 20" fixed aperture at 2700~\AA, see \citet{Kinney1993}. The integrated polarization degree and polarization angle are 11.4\% $\pm$ 0.2\% and 97.6$^\circ$ $\pm$ 0.4$^\circ$, respectively. These values are in good agreement with the WUPPE polarization measurement of NGC~1068 made by \citet{Code1993}, who found 12.9\% $\pm$ 1.9\% and 112$^\circ$ $\pm$ 3.8$^\circ$ for an aperture of 6" $\times$ 12". Through the 1" aperture of the Faint Object Spectrograph (FOS) centered on the peak of the continuum emission, \citet{Antonucci1994} measured a degree of polarization of 17.2\% $\pm$ 1.1\% and a position angle of 91.1$^\circ$ $\pm$ 1.8$^\circ$ in the range 2460–2760 \AA. By simulating a 1" diameter aperture on the FOC data, we obtained 10.7\% $\pm$ 0.1\% and a position angle of 91.0$^\circ$ $\pm$ 0.7$^\circ$(see Fig.~\ref{fig:comparison_A}).

\begin{figure}
    \centering
    \includegraphics[width=\linewidth]{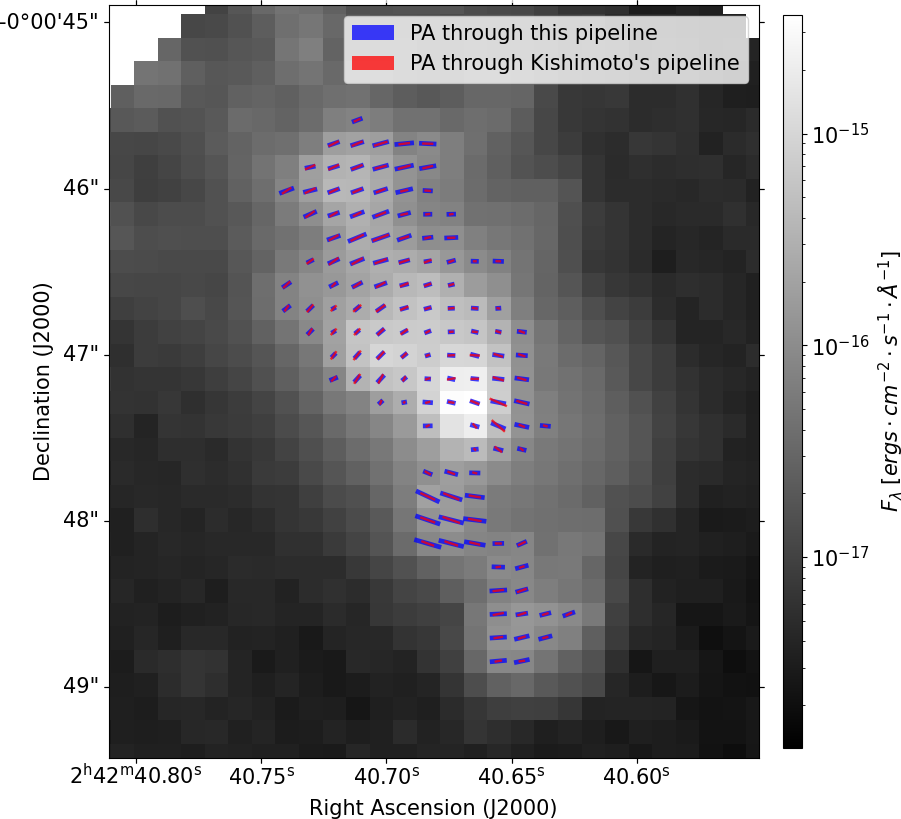}
    \caption{Direct comparison of the polarization map obtained by this pipeline to the one obtained by \citet{Kishimoto1999} for the same 10 pixel binning, without smoothing and taking the intersection of both maps cut at $\left[\text{S/N}\right]_P \geq 3$ and $\left[\text{S/N}\right]_I \geq 30$.}
    \label{fig:comparison_K}
\end{figure}
\begin{figure}
    \centering
    \includegraphics[width=\linewidth]{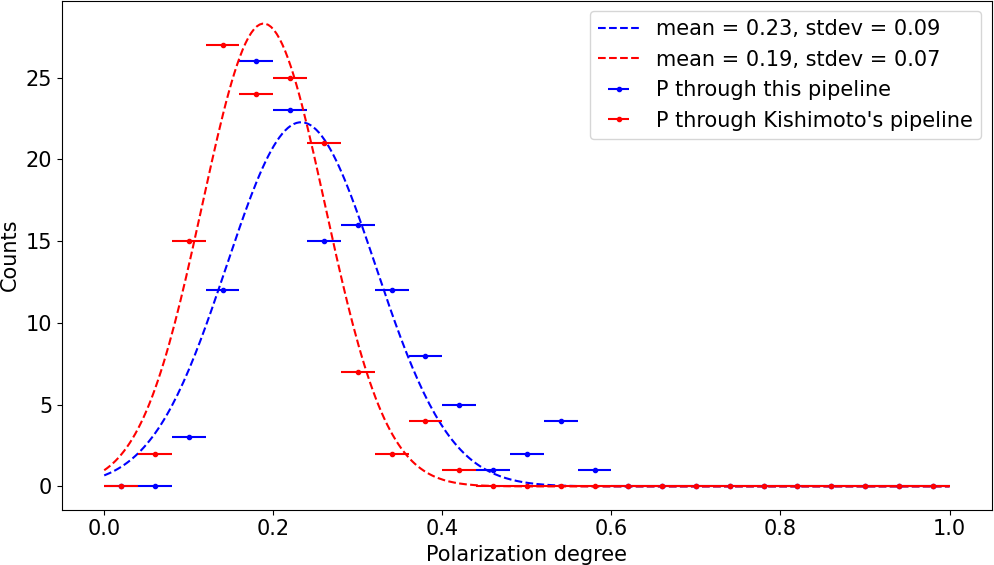}
    \caption{Distribution of the polarization degree obtained by this pipeline to the one obtained by \citet{Kishimoto1999} in the same cut at $\left[\text{S/N}\right]_P \geq 3$ and $\left[\text{S/N}\right]_I \geq 30$.}
    \label{fig:comparison_K_pd}
\end{figure}
\begin{figure}
    \centering
    \includegraphics[width=\linewidth]{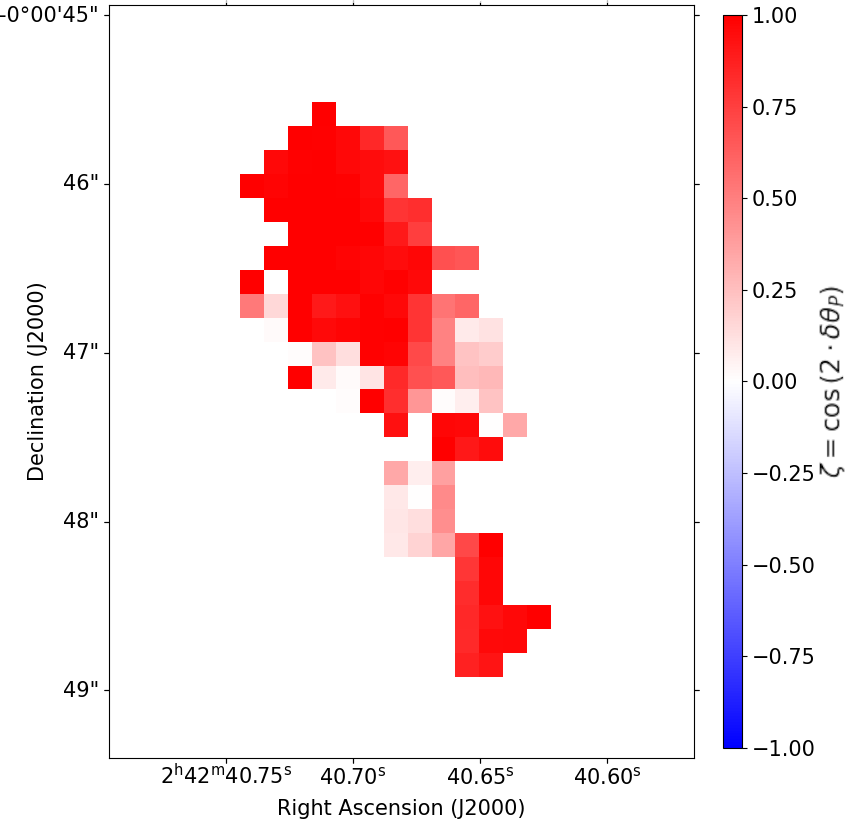}
    \caption{Degree of alignment of the polarization angle obtained by this pipeline to the one obtained by \citet{Kishimoto1999} in the same cut at $\left[\text{S/N}\right]_P \geq 3$ and $\left[\text{S/N}\right]_I \geq 30$.}
    \label{fig:comparison_K_pa}
\end{figure}

Another test was made by comparing our polarization map to the one of \citet{Kishimoto1999}. In this case, we used a 10 pixel binning, without smoothing, and taking the intersection of both maps cut at $\left[\text{S/N}\right]_P \geq 3$ and $\left[\text{S/N}\right]_I \geq 30$. The results are shown in Fig.~\ref{fig:comparison_K}. The polarization pattern reproduces the shape of the figure from \citet{Kishimoto1999}, as the exact same location of the statistically significant pixels. We note, however, a slight difference in the vector length, likely due to a different background estimation. In \citet{Kishimoto1999}, the background is estimated by finding a plateau in the outskirt of the radial flux profile as a function of the distance from the center, this plateau value is taken to be the image background. As this sort of technique fails to properly estimate a background value for polluted sources (as we will see with IC~5063 in section \ref{IC5063}), we chose to implement a more general approach, hence the possible difference.

In Fig.~\ref{fig:comparison_K_pd}, we compare the measured polarization degree from each analysis pipeline using a Gaussian fit. As observed on the superimposed polarization maps, the detected polarization degree is slightly higher from this pipeline with a mean at 23\% and a larger distribution than from \citet{Kishimoto1999}, which shows a mean at 19\% and a more peaked distribution. Hence, the two pipelines give similar results within the uncertainties. The difference most likely comes from the method of background subtraction and how we estimate the uncertainties, as we use the debiased polarization degree.

To compare the alignment of the obtained polarization angles in each pixel, we use circular statistics and introduce the metric $\zeta$ \citep{CH2019}, as defined in Eq.~\ref{eq:zeta}, with $\theta_1$ as the polarization angle from \citet{Kishimoto1999} and $\theta_2$ the one from this pipeline. 

\begin{gather}
    \zeta = \cos{(2\delta\theta)}\label{eq:zeta},\\
    \nonumber \text{with } \delta\theta = \frac{1}{2}\arctan{\left[\frac{\sin{(2\theta_1)}\cos{(2\theta_2)}-\cos{(2\theta_1)}\sin{(2\theta_2)}}{\cos{(2\theta_1)}\cos{(2\theta_2)}+\sin{(2\theta_1)}\sin{(2\theta_2)}}\right],}\\
    \nonumber \text{and } \theta_1, \theta_2 \text{ the angles to be compared.}
\end{gather}

Here, $\zeta$ is defined on the range of [$-1$, 1], such that two perfectly aligned distributions will have $\zeta$ = 1, two perpendicular distributions will have $\zeta$ = $-1$, and two distributions with no statistical alignment will have $\zeta$ = 0. 

\begin{figure}
    \centering
    \includegraphics[width=\linewidth]{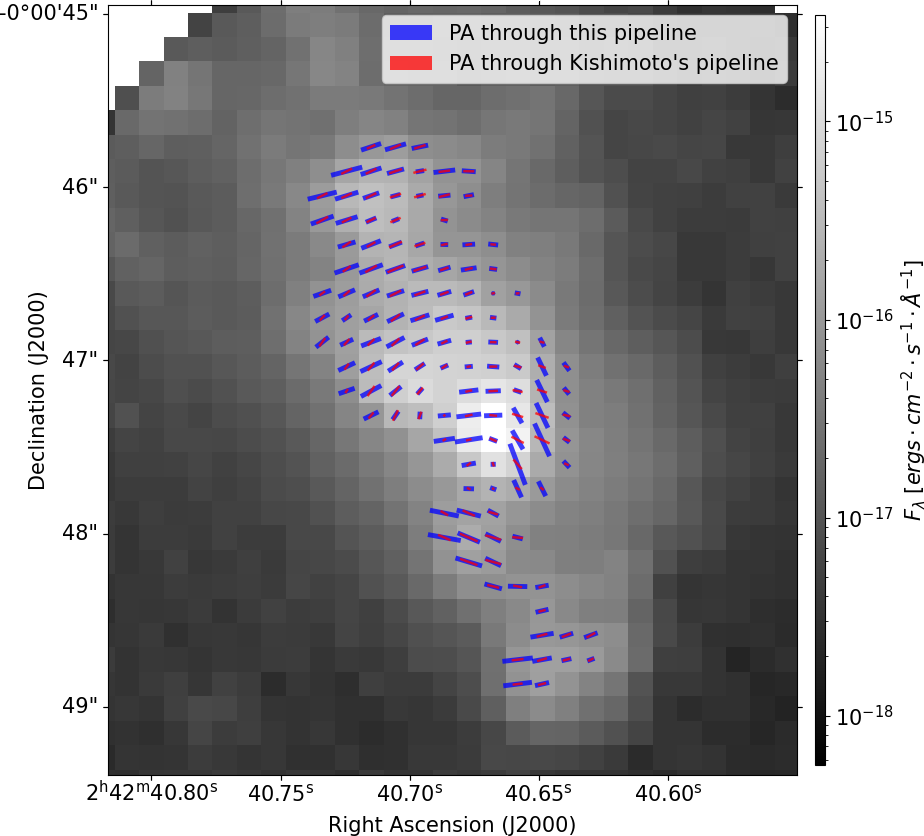}
    \caption{Same as Fig.~\ref{fig:comparison_K}, except that the alignment uses the shift values from \citet{Hodge1996}; i.e., there is no cross-correlation, highlighting the fact that the 2005's re-calibrated dataset have updated geometric properties in comparison to the dataset used in \citet{Capetti1995a,Capetti1995b} and \citet{Kishimoto1999}.}
    \label{fig:comparison_K_na}
\end{figure}

From Fig.~\ref{fig:comparison_K_pa}, we can see that both pipeline get almost identical polarization angles, $\zeta>0.8$, where the polarized flux is stronger and where the statistics are better. Outside of these regions of high S/N, and where background estimation and uncertainties becomes non-negligible, the alignment is less strong, but still in agreement. Since the whole HST/FOC dataset was re-processed in 2005 with new geometric references and the latest filter calibrations \citep{Kamp2006}, the archival data are intrinsically different from the science data that were used during the FOC lifetime. This can explain the differences in the polarization degree histogram in Fig.~\ref{fig:comparison_K_pd} and the polarization angle misalignment in the clouds N-W and S-E of the nucleus location in Fig.~\ref{fig:comparison_K_pa}. Additionally, it became important to align the different observation by cross-correlation and not by using the images shifts provided by \citet{Hodge1996}. These were used for the data reduction by \citet{Kishimoto1999} but the new best geometric reference modified these values, as it can be seen in Fig.~\ref{fig:comparison_K_na}, in which the observations are incorrectly shifted using the pixel shifts values from \citet{Hodge1996}.

\begin{figure*}
    \centering
    \includegraphics[width=\textwidth]{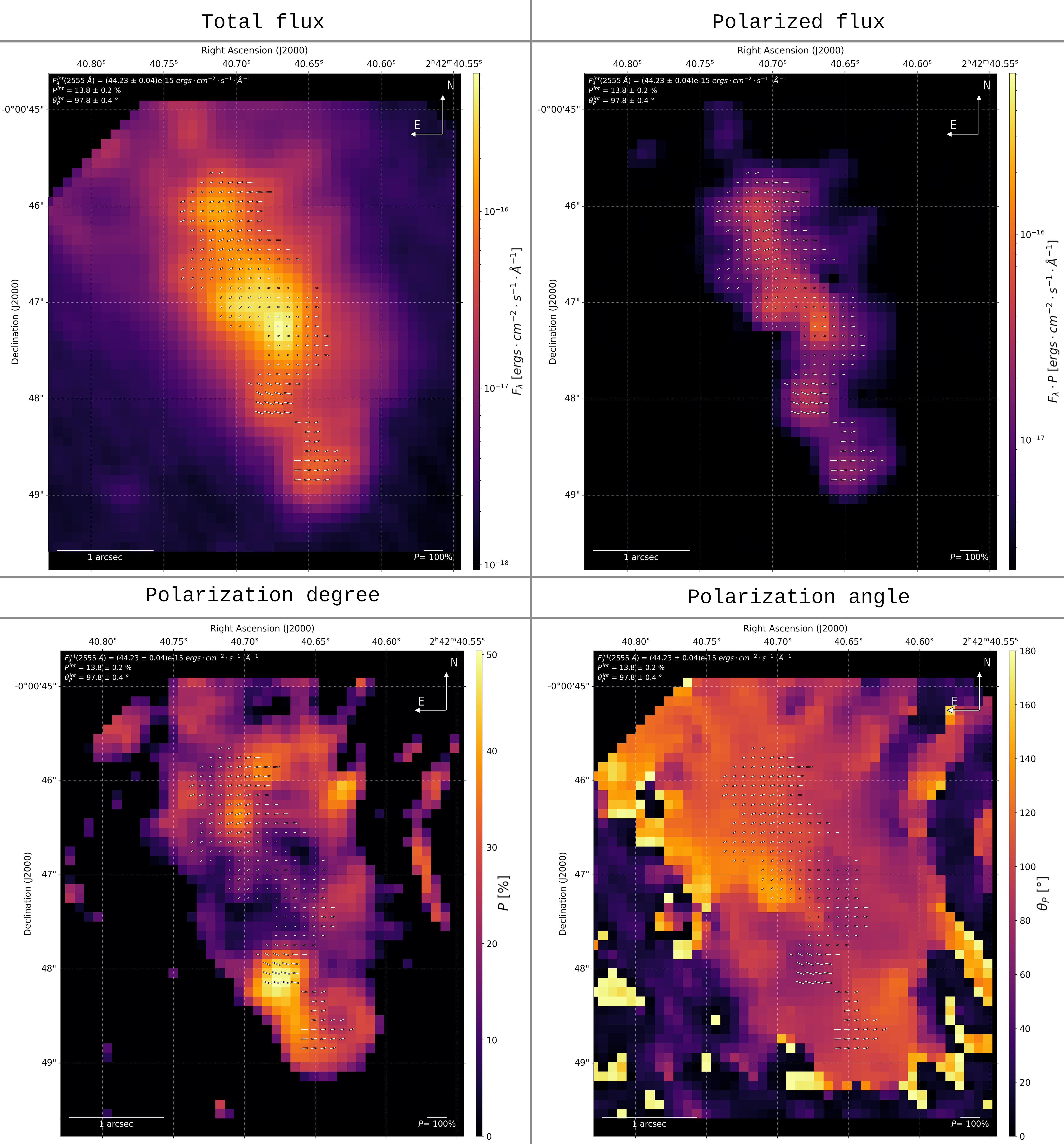}
    \caption{Four different zoomed-in outputs of the pipeline for NGC~1068 with the polarization map superimposed. The polarization vectors are only displayed for the selected cut of $\left[\text{S/N}\right]_P \geq 3$ and $\left[\text{S/N}\right]_I \geq 30$. The integrated values are computed on the full FoV ($P^{int}$ and $\theta_P^{int}$). \textit{Top-left}: Total flux $F_\lambda$ (erg.s$^{-1}$.cm$^{-2}$.\AA$^{-1}$, in log-scale). \textit{Top-right}: Polarized flux $F_\lambda \cdot P$ (erg.s$^{-1}$.cm$^{-2}$.\AA$^{-1}$). \textit{Bottom-left}: Polarization degree $P$ (\%). \textit{Bottom-right}: Polarization angle $\theta_P$ (in °, taken in the trigonometric direction with north being 0°).}
    \label{fig:NGC1068_f_pf_p_pa}
\end{figure*}

A deeper examination of the flux and polarization pattern of NGC~1068 is presented in the mosaic of Fig.~\ref{fig:NGC1068_f_pf_p_pa}. Here we plot zoomed-in maps of the AGN, showing the total flux (top-left), polarized flux (top-right), polarization degree (bottom-left), and polarization position angle (bottom-right). Comparing the total and polarized flux map, we see that the sinuous shape of the winds is much more emphasized by the polarized flux, where reprocessing is clearly revealed. This makes it possible to identify the geometry of the winds with greater precision, using the fact that polarization offers a better contrast than total flux images.

\begin{figure*}
    \centering
    \includegraphics[width=\textwidth]{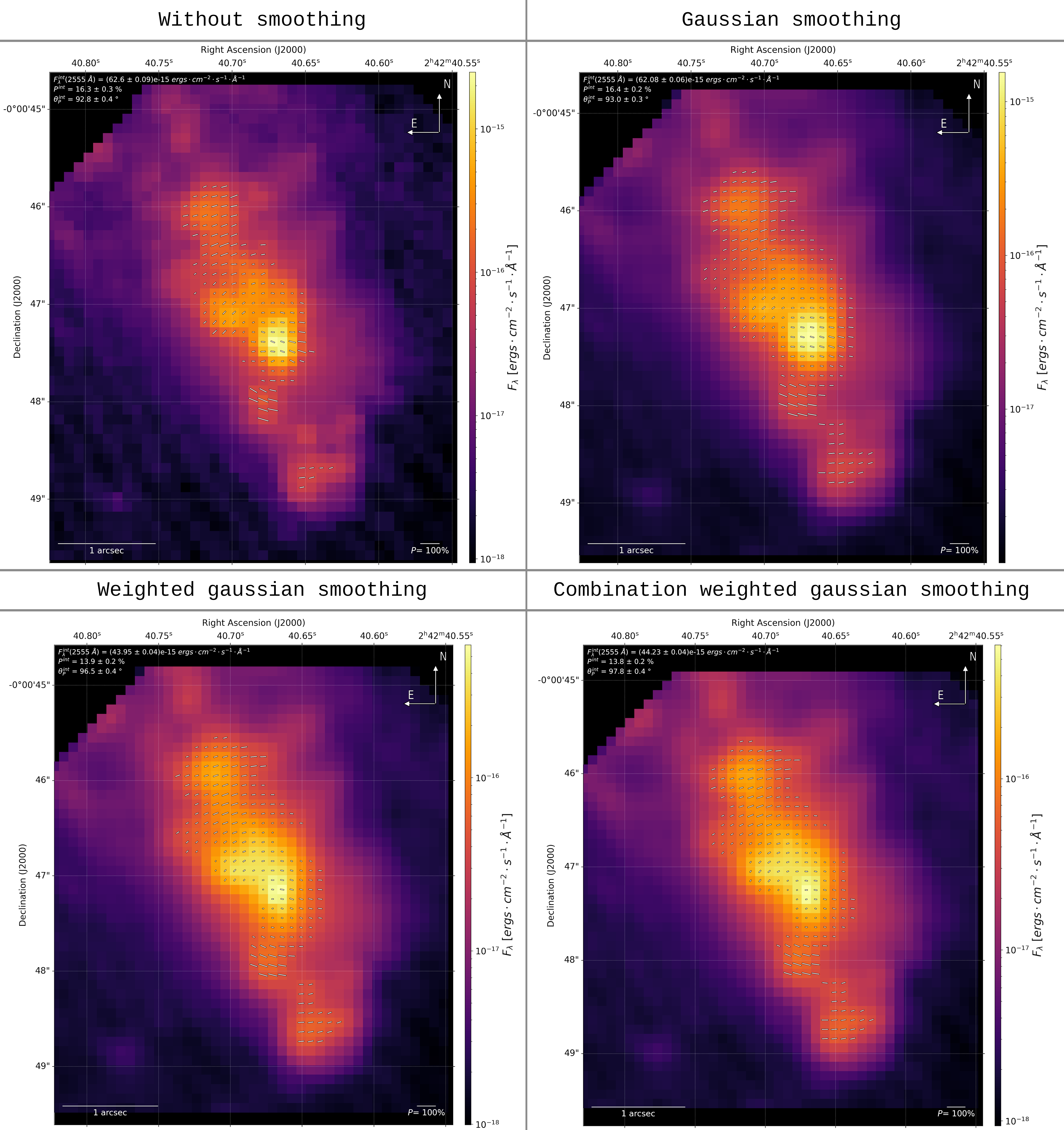}
    \caption{Four different zoomed-in outputs of the pipeline for NGC~1068 with the polarization map superimposed. The polarization vectors are only displayed for the selected cut of $\left[\text{S/N}\right]_P \geq 3$ and $\left[\text{S/N}\right]_I \geq 30$. This juxtaposition shows different smoothing methods, all maps are binned to 0.1" pixels. \textit{Top-left}: Without smoothing. \textit{Top-right}: Simple Gaussian smoothing of FWHM of 0.2". \textit{Bottom-left}: Gaussian smoothing with FWHM of 0.2" where pixels are weighted with their inverse squared error. \textit{Bottom-right}: Combination and weighted Gaussian smoothing where the different observation through a same polarizer filter are both averaged and smoothed at the same time for a better reduction.}
    \label{fig:smoothing_NGC1068}
\end{figure*}

Different types of data smoothing were implemented in the pipeline to allow reproduction of older data reductions and results. A simple Gaussian smoothing with a FWHM determined by a pixel radius allow to reproduce the data reduction as usually done by previous papers \citep[see][]{Capetti1995b,Antonucci1994}. However, for future works, we prefer to use smoothing weighted on each pixel related error. A comparison between these smoothing methods can be seen in Fig.~\ref{fig:smoothing_NGC1068} and it highlights the better S/N permitted by a simple Gaussian smoothing and the improved definition of spatially resolved structures using weighted methods.

\section{Uncharted FOC observation of IC~5063}
\label{IC5063}
Once we have made sure our pipeline produces valid polarization maps, we can undertake the exploration of one of the unpublished AGN observations in the FOC archives: IC~5063. We leave the remaining unexplored AGNs for the next papers of this series. 

IC~5063 is a nearby elliptical galaxy ($z \approx$ 0.01135, corresponding to a Hubble distance of $\sim$ 48.32~Mpc in the standard $\Lambda$CDM model). It is a radio-loud galaxy, with a bright red nucleus. The latter characteristic can either come from a very steep non-thermal spectrum (with spectral index -4.5) or by re-radiation from hot dust with a black-body color temperature of 650 K \citep{Axon1982}. In 1987, a high polarization degree (17.4 $\pm$ 1.3 \% at PA $\approx 4 \pm 5$° in $H$ and $K$ bands) was measured in near-infrared for a $2.25$ arcseconds aperture centered on the nucleus \citep{Hough1987}. This suggests a non-thermal synchrotron source for the near-IR emission from the nucleus. The detection of a strong, broad H$\alpha$ emission seen in polarized flux also suggests the existence of a hidden broad-line region \citep{Inglis1993}, pointing towards a hidden type-1 AGN nested in the dusty heart of an elliptical galaxy. Indeed, a prominent dust lane has been observed along the long axis of IC~5063, mostly concentrated on the northern side. The symmetrical distribution of this dust lane and its continuity outside of the nucleus spanning a few kiloparsecs suggest an external origin, most probably a previous merger, as such structures are unlikely to survive many dynamical timescales \citep{Colina1991}. Finally, it has been observed that this AGN displays strong interactions between the ISM and its radio jets \citep[radio position angle $\sim$ 115$^\circ$,][]{Morganti1998} that introduce complex emission regions along the jets \citep{Oosterloo2000}. These regions spanning a few hundred parsecs are a perfect observational target for the FOC thanks to its fine spatial resolution (at $z =$ 0.01135, 1" equals to 241.7~pc). 

IC~5063 was observed by the FOC on Feb 25, 1998 (program ID 5918). The observation used the F502M filter centered around 4985~\AA, for an exposure time of 5\,261 seconds through each POL0, POL60 and POL120 filters. This adds up to a total observation time of 15\,783 seconds. The observation was reduced in total flux by \citet{Dasyra2015}, in particular, their Fig.~8, and allowed for the identification of discrete gas-outflow starting points along the radio jets. However, no polarization study has ever been published despite the rather good quality of the data.

Some concern may arise about the potential contamination (dilution) from the extended [O~III] polar emission that could impact our resulting maps. Indeed, we can see from \citet{Venturi2021} Fig.~1 (c) that the HST/FOC FoV is totally embedded in the [O~III] emission region. These emission lines are formed when atoms from the polar region are photo-ionized by the continuum radiation from the central source. Photo-ionization produces unpolarized photons \citep{LeeBlandfordWestern1994,Lee1994}, so our whole map is subject to polarization dilution by the [O~III] emission. As such, it likely decreases the observed polarization degree ($P$), but it should not change the polarization angle ($\theta_P$). There should not be misleading polarization patterns.

\subsection{Characteristics of the optical polarization map}
\begin{figure*}
    \centering
    \includegraphics[width=\textwidth]{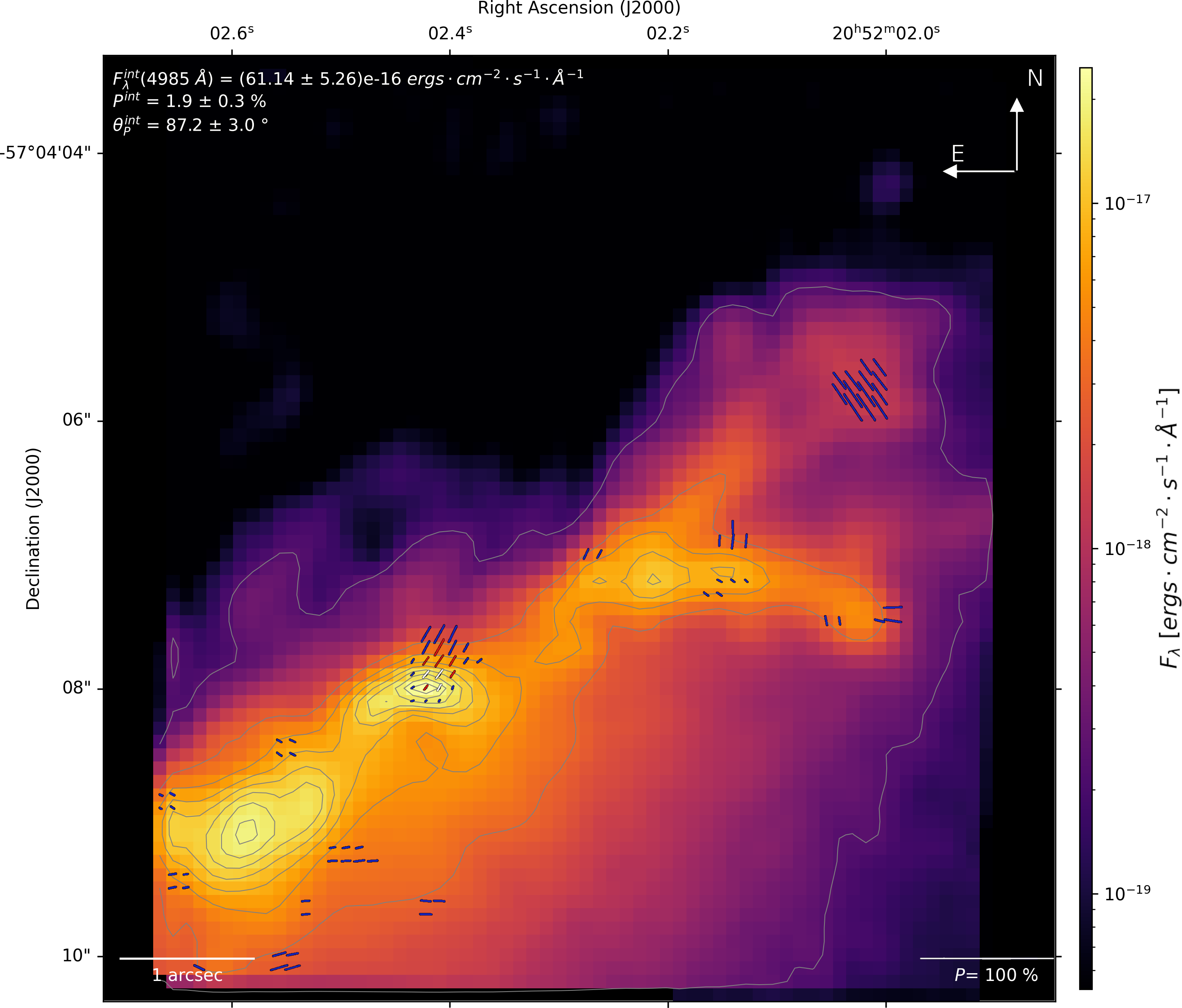}
    \caption{Total flux F$_\lambda$ (erg.s$^{-1}$.cm$^{-2}$.\AA$^{-1}$, color-coded) of IC~5063, with the polarization information superimposed to the image using white vectors. The data is re-sampled to have bins of 0.1" and smoothed with a Gaussian kernel of FWHM of 0.2". The polarization vectors are displayed for $\left[\text{S/N}\right]_I \geq$ 30 and $\left[\text{S/N}\right]_P \geq$ 3 in white, for $\left[\text{S/N}\right]_P \geq$ 2 in red and for $\left[\text{S/N}\right]_P \geq$ 1 in blue. Their lengths are proportional to $P$. The contours are displayed from 1\% to 99\% every 10\% of the maximum value of $1.727 \cdot 10^{-15}$ erg.s$^{-1}$.cm$^{-2}$.\AA$^{-1}$.}
    \label{fig:IC5063}
\end{figure*}

\begin{figure*}
    \centering
    \includegraphics[width=\textwidth]{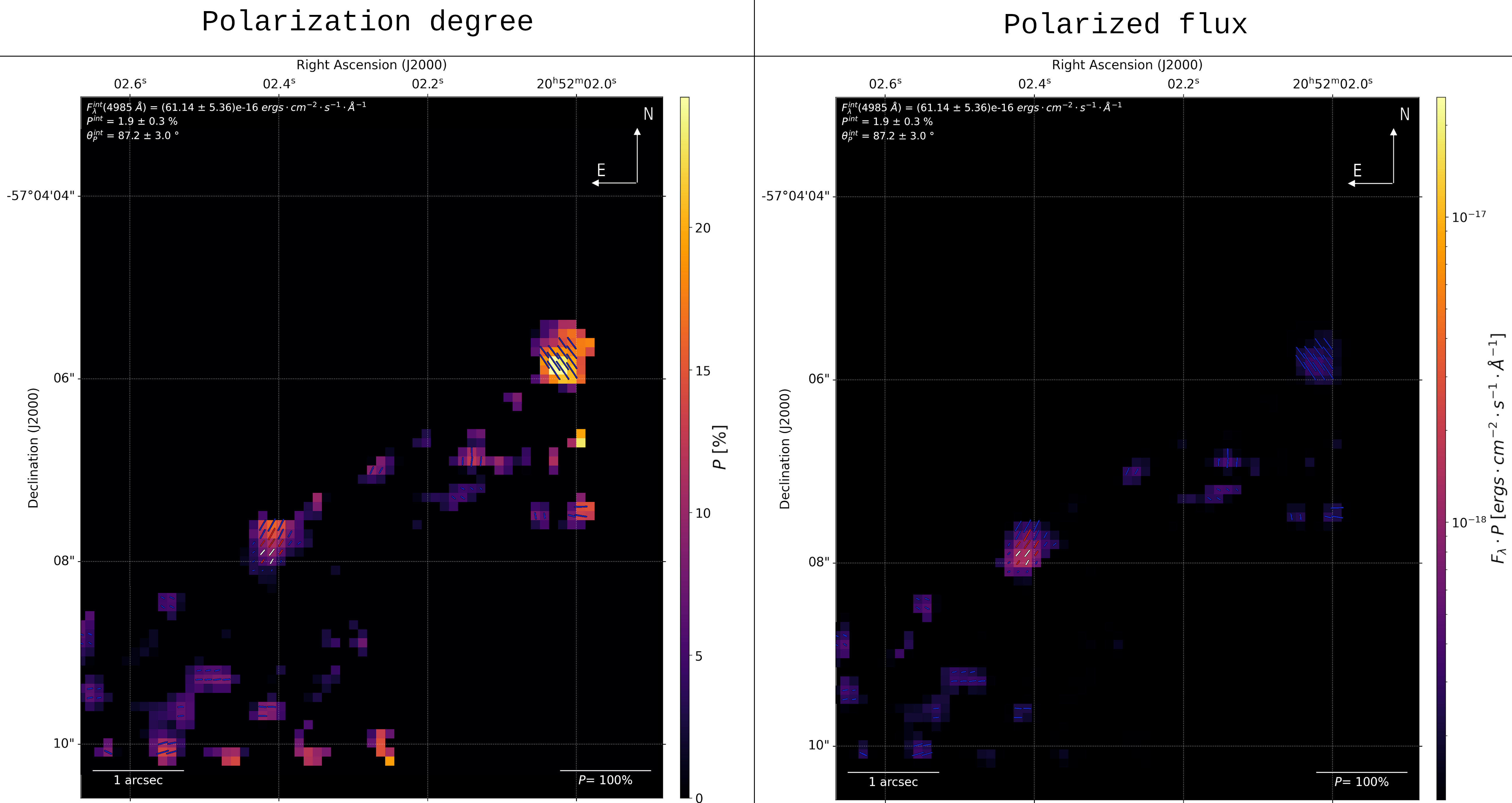}
    \caption{Two different outputs of the pipeline for IC~5063 with the polarization map superimposed. The polarization vectors are only displayed for the selected cut of $\left[\text{S/N}\right]_I \geq$ 30 and $\left[\text{S/N}\right]_P \geq$ 3 in white, for $\left[\text{S/N}\right]_P \geq$ 2 in red and for $\left[\text{S/N}\right]_P \geq$ 1 in blue. The integrated values are computed on the full FoV ($P^{int}$ and $\theta_P^{int}$). \textit{Left}: Polarization degree $P$ (\%). \textit{Right}: Polarized flux $F_\lambda \cdot P$ (erg.s$^{-1}$.cm$^{-2}$.\AA$^{-1}$).}
    \label{fig:IC5063_f_pf_p_pa}
\end{figure*}

We processed the observation of IC~5063 through our pipeline and we present the resulting 4985~\AA-centered polarization map in Figs.~\ref{fig:IC5063} and \ref{fig:IC5063_f_pf_p_pa}. We show the total intensity map with the polarization information superimposed to it. We rebinned the data to get individual spatial bins of 0.1" and the pixels were smoothed by a Gaussian kernel of standard deviation 0.2". Only the bins with a $\left[\text{S/N}\right]_I \geq 30$ and $\left[\text{S/N}\right]_P \geq 3$ have their polarization vectors shown. We lowered the cut to $3\sigma$ in polarization degree due to lower polarized flux coming from the the AGN compared to the $5\sigma$ cut on the observation of NGC~1068.

The total flux image is dominated by a croissant-shaped region that is situated near the base of the jets (RA 20h52m02.4s, DEC -57°04'08"), but it does not necessarily corresponds to the location of the hidden nucleus. We remind the reader that the central engine is obscured by a circumnuclear reservoir of dust, so it is not directly visible in total flux. This bright croissant most likely results from: a) the re-emission of the ISM by the interaction of the jet with the host galaxy and/or b) the scattering of photons thermally emitted from the central accretion disk. The opposite jets, invisible in UV/optical light due to Doppler beaming outside our line-of-sight, extend towards the southeast and northwest directions. The lobes of the jets, detected in radio and seen superimposed on the polarization map in Fig.~\ref{fig:IC5063_radio}, match two regions with a higher intensity than the background light, together with specific morphologies. The northwest lobe region shows a peculiar $V$ shape that could result from the interaction of the northern jet with the material expelled by the central engine before the onset of jet activity. This material is likely to be the narrow line region (NLR) across the host galaxy that we observe in many radio-quiet AGNs (see the case of NGC~1068). In the southern part, a less intense, diffuse clumpy emission can be seen. This can be due to the fact that the southeast jet and lobe are obscured by the dusty disc of the host galaxy and the medium they interact with is responsible for the observed optical/UV radiation.

\subsection{Region-by-region exploration and correlation to radio and IR data}
\begin{figure}
    \centering
    \includegraphics[width=\linewidth]{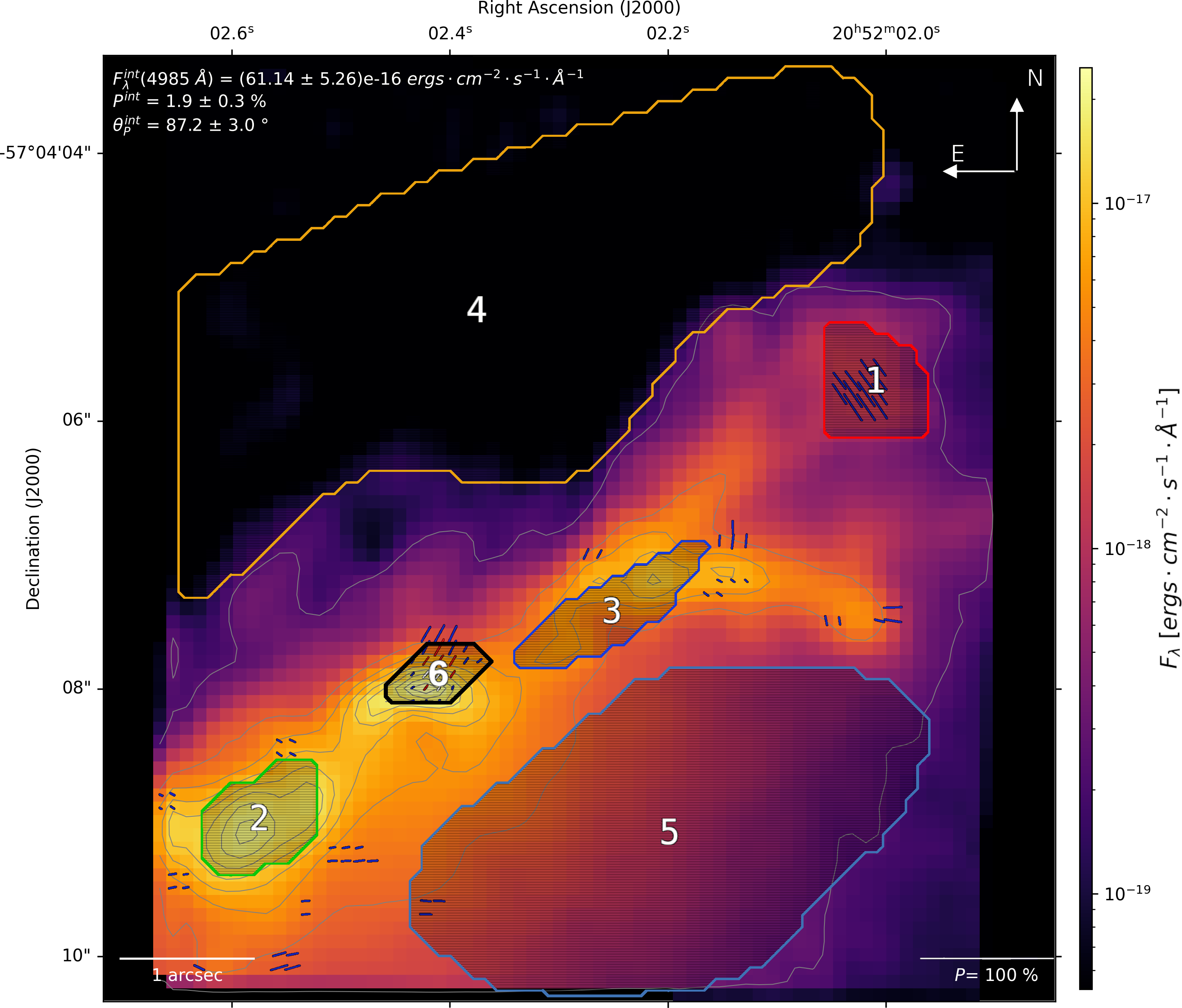}
    \caption{Optical polarization map for IC~5063 obtained through the pipeline. The polarization vectors are displayed for $\left[\text{S/N}\right]_I \geq$ 30 and $\left[\text{S/N}\right]_P \geq$ 3 in white, for $\left[\text{S/N}\right]_P \geq$ 2 in red and for $\left[\text{S/N}\right]_P \geq$ 1 in blue and the overlaid regions are listed in Tab.~\ref{tab:IC5063_int_regions} and detailed in the text.}
    \label{fig:IC5063_map_regions}
\end{figure}

\begin{figure}
    \centering
    \includegraphics[width=\linewidth]{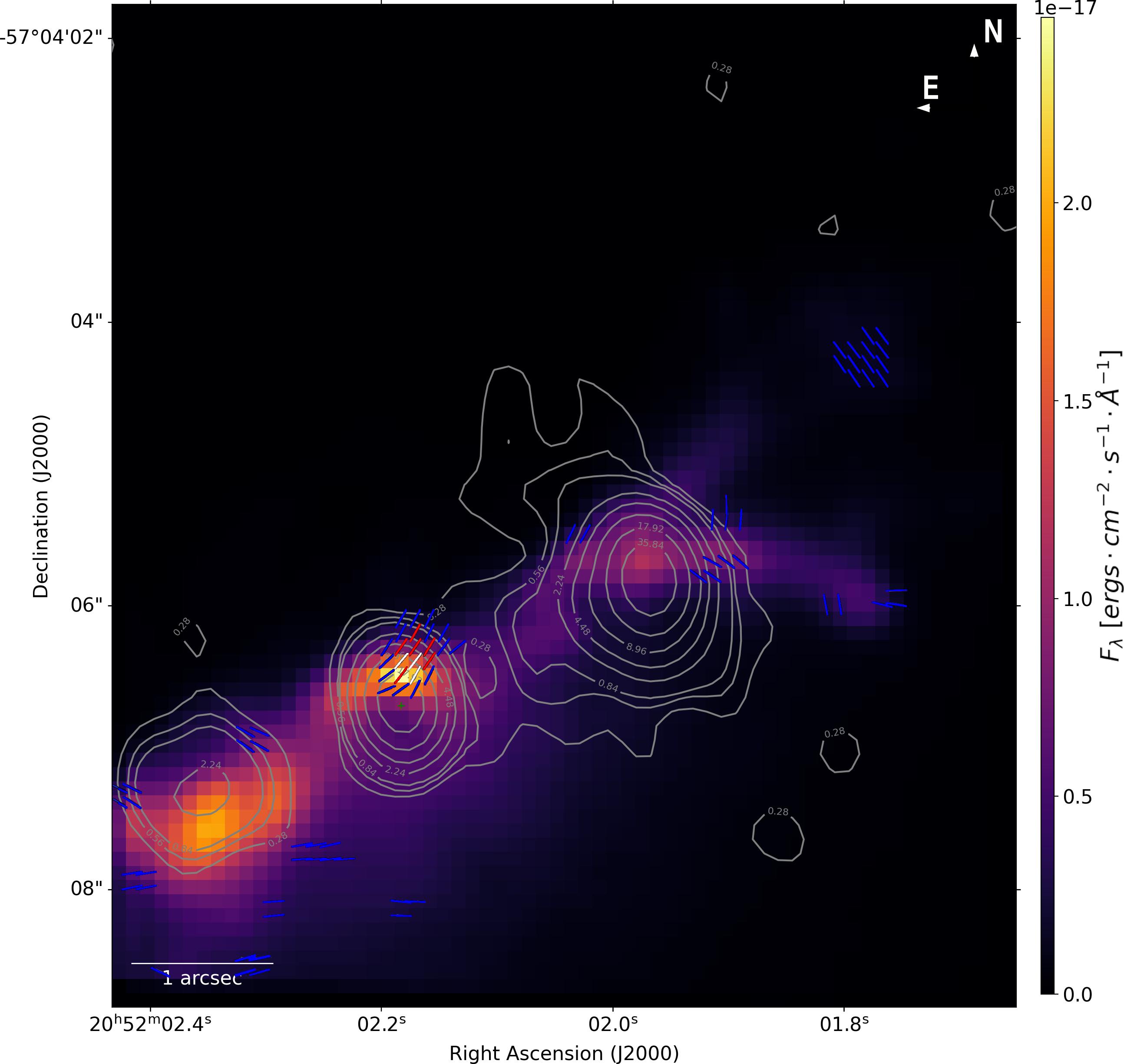}
    \caption{18~GHz ATCA map from \citet{Morganti1998} superimposed onto our HST/FOC map. Both maps have been aligned on the supposed nucleus location. The optical polarization vector have a fixed length to better highlight their orientation.}
    \label{fig:IC5063_radio}
\end{figure}

\begin{figure}
    \centering
    \includegraphics[width=\linewidth]{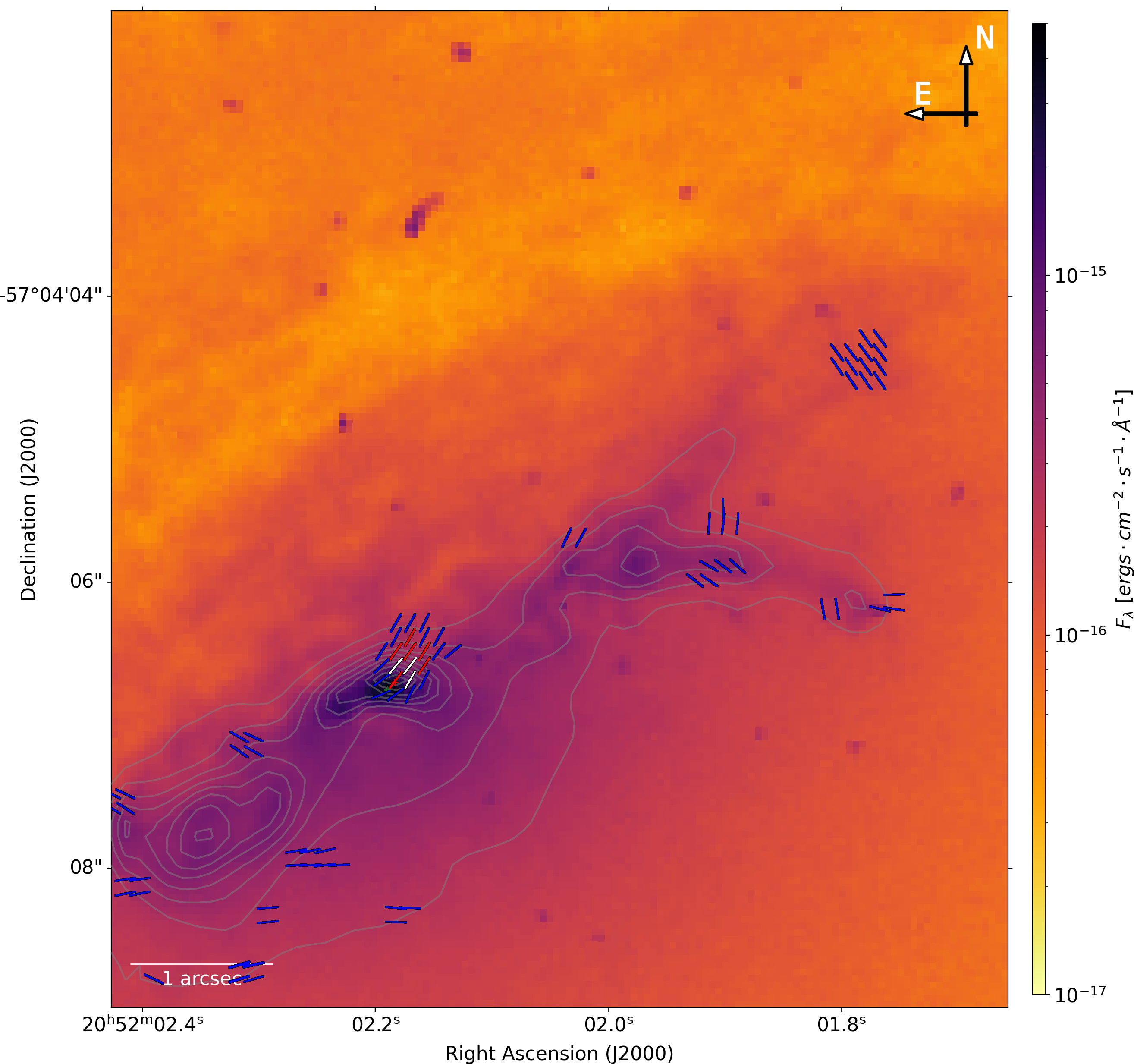}
    \caption{Total flux of IC~5063 at $5997$ \AA\ observed by HST/WFPC2 in 1995. The superimposed polarization vectors, shown with a constant length for better visualization, are taken from the polarization map at $4985$ \AA\ from this paper. They are displayed for the selected cut of $\left[\text{S/N}\right]_I \geq$ 30 and $\left[\text{S/N}\right]_P \geq$ 3 in white, for $\left[\text{S/N}\right]_P \geq$ 2 in  red and for $\left[\text{S/N}\right]_P \geq$ 1 in blue (less significative). Both maps have been aligned on the croissant-shaped high flux region.}
    \label{fig:IC5063_ir}
\end{figure}

IC~5063 appears to show a lot of complex interactions between the jets, the AGN winds and the local galactic material. Thus, we divided the HST/FOC map in six distinct regions. These regions are listed in Tab.~\ref{tab:IC5063_int_regions} and delimited in Fig.~\ref{fig:IC5063_map_regions}. These specific regions are individually explored through the polarization information made available by the HST/FOC observation and correlated with observations at other wavelengths:
\begin{itemize}
    \item In Fig.~\ref{fig:IC5063_radio}, we show the superposition of our HST/FOC map with the 18~GHz ATCA map provided by \citet{Morganti1998}. Due to small discrepancies between between ATCA and HST respective WCS coordinates ($\sim$ 0.5" error), both maps were manually aligned with the hypothesis that the hidden nucleus is right below the croissant-shaped high flux and is the center of the radio emission. 
    \item In Fig.~\ref{fig:IC5063_ir}, we make use of the F606W filter of the Wide Field and Planetary Camera 2 (WFPC2) of the HST to highlight the presence and position of the dust lane. Due to differences of the HST pointing between 1995 and 1998 ($\sim$ 1" error), the WFPC2 and FOC maps were manually aligned with the hypothesis that the croissant-shaped high flux in both near-UV and near-IR are of similar origin. We inverted the color-map for the near-IR map in order to better see the extension of the dust-lane on the northern part of the AGN. 
\end{itemize}

\begin{table}
    \centering
    \resizebox{\linewidth}{!}{%
        \begin{tabular}{c|c|c|c|c}
            Region  & $F_{\lambda}^{int}$   & $P^{int}$         & $\theta_P^{int}$  & radio PA - $\theta_P^{int}$  \\ 
            \hline
            FoV     & $611.4 \pm 52.6$      & $1.9 \pm 0.3$     & $87.2 \pm 3.0$    & $\sim$ 28 \\
            1       & $5.684 \pm 0.085$     & $20.4 \pm 2.3$    & $38.5 \pm 3.4$    & $\sim$ 77 \\
            2       & $87.02 \pm 0.18$      & $0.7 \pm 0.4$     & $36.8 \pm 13.7$   & $\sim$ 78 \\
            3       & $39.16 \pm 0.13$      & $0.7 \pm 0.6$     & $9.9 \pm 21.9$    & $\sim$ 105 \\
            4       & $4.247 \pm 0.338$     & $29.4 \pm 13.2$   & $88.4 \pm 11.4$   & $\sim$ 26 \\ 
            5       & $90.38 \pm 0.32$      & $4.9 \pm 0.6$     & $91.0 \pm 3.1$    & $\sim$ 23 \\     
            6       & $26.75 \pm 0.11$      & $5.8 \pm 0.7$     & $142.4 \pm 3.6$   & $\sim$ -27
        \end{tabular}%
    }
    \caption{IC~5063 optical polarization maps integrated over the specific regions displayed on Fig.~\ref{fig:IC5063_map_regions}. The first line corresponds to the full FoV integrated data. Fluxes are in 10$^{-17}$~ergs cm$^{-2}$ s$^{-1}$ \AA$^{-1}$, $P^{int}$ in percents and $\theta_P^{int}$ in degrees. The radio position angle (PA) has been measured by \citet{Morganti1998} and is about 115$^\circ$.}\vspace{-15pt}
    \label{tab:IC5063_int_regions}
\end{table}

\subsubsection{Region 1: Northern AGN winds}
The first region we investigate corresponds to a zone that is beyond the interaction of the northern jet and the local medium. It is relatively unperturbed and matches the location of the extended NLR seen in [O III] emission by \citet{Colina1991}. The polarization degree we measure is high, on the order of 36\%, associated with a polarization position angle of $\sim$ 30$^\circ$. Compared to the position angle of the extended [O III] emission and of the radio jets (115$^\circ$), the polarization angle appears perpendicular. This, together with the high degree of polarization, indicates that scattering prevails in this region. The light we observe directly comes from the obscured central engine through perpendicular scattering onto the winds that act like a mirror. In addition, this result agrees with the polarization angle and degree observed in optical by \citet{Inglis1993} ($\sim$ 10--30\% at $\sim$ 30--50$^\circ$). Any spectropolarimetric attempt to measure broaden Balmer lines in polarized flux should then focus on this region to be free from the turbulence and emission caused by the northern jet and lobe. Interestingly, our independent detection of the undisturbed NLR allows us to say that IC~5063 did have polar winds before the onset of jet activity, but the ionization cone and the NLR structures have been almost completely wiped out by the jet within the first arc-seconds around the core. 

\subsubsection{Region 2:  Southern lobe}
The second brightest spot in flux at the 4985~\AA-centered optical waveband is situated on the southwestern part of the AGN and corresponds to a zone where the counter jet seem to interact with the local medium. This region of high flux is almost completely depolarized. We can see from the superposition of the radio map to the optical data (Fig.~\ref{fig:IC5063_radio}) that this depolarized region also corresponds to the southern lobe. This depolarization seen in near-UV may originate from a scattering medium perturbed by the jet and lobe, resulting in a complex reprocessing environment producing polarized photons with various polarization angles and, thus, resulting in a net depolarization of the photon flux. We note, however, the presence of a circular polarization pattern where the lobe encounters the ISM. In these region, the medium might be kinetically aligned by the jet's lobe pushing the medium out of its way, but the statistical significance of the detection per pixel is below 3$\sigma$. However, Table~\ref{tab:IC5063_int_regions} shows that integrating the whole lobe region gives a polarization $\sim$1\% with a $\left[\text{S/N}\right]_P$ of 6.5. This is a clear detection of a polarized source at a PA $\sim 15^{\circ}$, almost perpendicular to the jet direction. 

\subsubsection{Region 3: Northern lobe}
From \citet{Morganti1998}, we know that this AGN is characterized by fast gas outflows fueled from the central core. In this region located between the hidden nucleus and the unperturbed wind, where the radio map highlight the presence of the northern lobe, we detected a low degree of polarization ($\sim2.6\pm0.5$\%), probably for the same reason as stated above for the southern lobe. Also, similarly to the southern lobe, the associated PA $\sim 10^{\circ}$ corresponds to a direction that is perpendicular to the radio jet. Furthermore, the V-shape seen in the 4985~\AA--centered optical waveband most likely indicate a highly interactive region where the radio jet, the polar wind, and the ISM meet.

\subsubsection{Region 4: Dust lane}
On the northern part of the HST/FOC map, we can see a clear cut in logarithmic scale on the flux intensity map. We know from \citet{Morganti1998}, in particular, their Fig.~4, that this obscured region in optical corresponds to the extension of the dust lane, which is confirmed by the WFPC2 infrared map (Fig.~\ref{fig:IC5063_ir}). Using the spatial capabilities of the FOC, we can isolate and integrate the polarization from this dusty region. Given the low flux from this specific region, we get higher errors but we do observe a high polarization degree, around 18\% at $\sim$ 84$^\circ$, exactly along the position angle of the dust lane. The high polarization degree and the polarization angle parallel to the dust lane strongly suggest dichroic absorption from starlight in the foreground dust lane. The integrated PA could then highlight the large-scale ordered magnetic field in the dust lane. 

\subsubsection{Region 5:  Diffuse medium}
The map of IC~5063 can essentially be divided in three parts : the northern section that is dominated by the dust lane, the southeast to northwest diagonal that corresponds to the highly asymmetric AGN, and the southern part that is essentially the ISM, a diffuse medium where \citet{Maksym2020} identified "dark rays." That is to say, the projected shadow of the circumnuclear dusty torus. For completeness, we investigated this region and found a low polarization degree ($\sim$ 1\%) and with a polarization position angle $\sim$ 123$^\circ$, namely, it is parallel to the jet radio axis. Because the southern region of the map is not as much obscured as the northern region, the observed polarization either results from dichroic absorption and re-emission of host starlight passing through the diffuse medium, or it could be the imprint of the magnetic field of the galaxy itself. Deeper (and longer) observations would be needed to assess the correct interpretation. 

\subsubsection{Region 6: Highly polarized knot}
This region is characterized by a high S/N associated to a strong polarized flux. It is just north of the estimated position of the nuclei and goes across the croissant-shaped region of highest flux. We measured a high polarization degree at $6.1\pm0.4$\% associated with a polarization angle of $151.2\pm1.7^{\circ}$. We learn from \citet{Lopez-Rodriguez2013} that the central 1.2 arcsec of IC~5063 in J, H, and K$_n$ band was measured to be 2.0±0.7\%, 2.5±0.9\%, and 7.8±0.5\%, respectively, and that the PA of polarization is wavelength-independent (within the error bars) and measured to be $3\pm6^\circ$ in the three filters. In this HST/FOC observation we get a high polarization degree because we are less diluted by the host starlight in the UV than in the IR, but the different PA we get with respect to \citet{Lopez-Rodriguez2013} indicates some pollution by either the jet or by the dust lane. As the PA-radioPA is similar to region 4, we estimate that we get a mix of signals from the dust lane and additionally from a region situated further away from the torus height. We can confirm that this is not the jet polarization, as otherwise we would see it all along the jet structure (also, due to Doppler boosting outside of our line-of-sight, this would not be a reasonable explanation). It is more likely that the polarization observed in UV comes  from dichroic absorption from the dust lane of AGN core photons scattered in our line of sight by the jet base or the ionized wind base.

\section{Discussion}
\label{Discussion}

We went on to test our new reduction pipeline against NGC~1068 and applied it to IC~5063 with the results detailed above. Here, we discuss this process, along with two more points related to our IC~5063 analysis and to the use of a different background noise estimator than usual.

\subsection{Past polarimetry of IC~5063}
\label{Discussion:IC5063}
In contrary to, for instance, NGC~1068 or NGC~4151 \citep{Marin2018,Marin2020}, IC~5063 has not been extensively observed in optical polarimetry. This is due to the fact that its optical polarization is strongly affected by both interstellar polarization and dilution by the host galaxy, but also because of the presence of the dust lane that hides a significant fraction of the AGN. From the archives, we were able to retrieve four papers that present optical and/or near-infrared polarimetric measurements of IC~5063. 

\citet{Martin1983} was probably the first to measure the optical linear polarization of IC~5063 using Pockels cell polarimeters on the Steward observatory. The observations were made with a Corning 4-96 filter (3800 -- 5600~\AA) and a 4" aperture. The authors found a polarization degree of 1.28\% $\pm$ 0.14\% at a polarization position angle of 10.1$^\circ$ $\pm$ 3.2$^\circ$. Broadband optical and near-infrared polarization measurements were also undertaken by \citet{Hough1987} with the Hatfield Optical-IR polarimeter on the 3.9m Anglo-Australian Telescope. The authors noted that the polarization decreases from B to J and then rises towards longer wavelength, while the polarization angle is wavelength-independent. \citet{Inglis1993} were the first to provide spectropolarimetric data with the same telescope as \citet{Hough1987}, mounted with a spectrograph and a rotating Thomson CCD. Strong, broad H$\alpha$ emission was discovered in polarized flux. The polarization obtained by \citet{Inglis1993} between 4500 and 7000~\AA\ is approximately constant : $\sim$ 1.7\% at 3$^\circ$. Finally, \citet{Lopez-Rodriguez2013} used the infrared polarimeter built by the University of Hertfordshire for the Anglo-Australian Telescope and measured polarization in J, H and Kn bands at four different sized apertures: 1.2, 2.0, 3.0, and 4.0 arcsecs. They found a larger and aperture-dependent polarization degree than in the optical, with a constant polarization position angle.

\begin{figure}
    \centering
    \includegraphics[width=\linewidth]{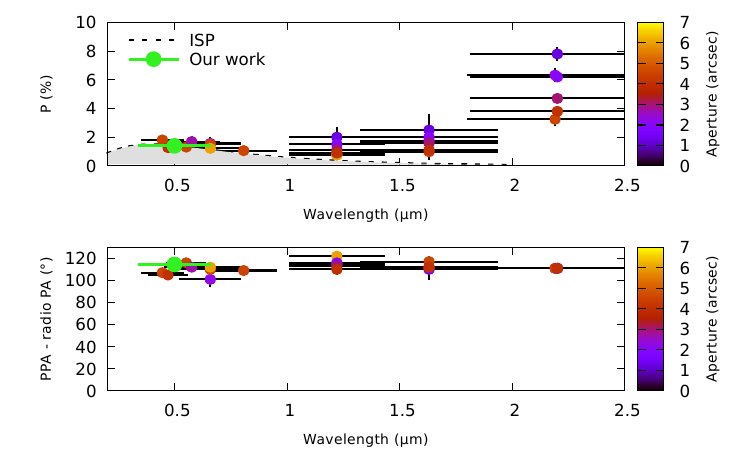}
    \caption{Compilation of published IC~5063 polarimetric measurements, prior to interstellar and starlight dilution correction. Top: Polarization degree. Bottom: Polarization angle minus the 
    radio position angle of the jet structure. The aperture used for each measurement is color-coded. The interstellar polarization (ISP) contribution is highlighted in gray. We simulated an aperture of radius 1 arcsec centered on the WCS reference point, situated on the NW part of our region 3, right at the beginning of the V-shape. We obtained a $1.0 \pm 0.4 \%$ polarization degree with an associated $7.9 \pm 10.5^\circ$ polarization angle. Our measurement is shown using a green cross. See text for details and references.}
    \label{fig:Comparison_archives}
\end{figure}

All archival polarimetric data are plotted in Fig.~\ref{fig:Comparison_archives}. We indicated the contribution of the interstellar polarization in gray, following the standard Serkowski's law \citep{Serkowski1975}. We used a synthetic 1.5" radius aperture on our IC~5063 map and checked whether we could reproduce the observation. The data we obtain is strongly position-dependent. Targeting the brightest spot in ultraviolet light, we only probe the jet base region, while displacing the aperture along the jet direction or placing it inside the dust lane region dramatically alter the observed polarization, as demonstrated in Sect.~\ref{IC5063}. Because we do not know what was the exact pointing of the various telescopes that obtained polarimetric data in the past, we made use of the fact that \citet{Inglis1993} measured both the narrow emission lines in the total flux and the broad emission line in polarized flux to estimate that the pointing encompassed the region of the map that is dominated by the polar winds (region 1 in Fig.~\ref{fig:IC5063_map_regions}). We thus extracted the continuum linear polarization from this region and obtained 1.3\% $\pm$ 0.3\% at 173.2$^\circ$ $\pm$ 4.3$^\circ$. This value is consistent with the previous measurement (see Fig.~\ref{fig:Comparison_archives}).

Playing with aperture values and position centers, we easily see that a variation of half an arcsecond can completely change the observed polarization in a source as complicated as IC~5063. Polarization imaging at a high resolution is thus needed to determine the polarization of each separate region, as integrating over a too large aperture would ultimately result in a mix of several emission and/or scattering mechanisms -- thus resulting in an erroneous explanation.

\subsection{Background noise estimation}
\label{Discussion:noise}
Estimating the background for point-like sources is rather simple: a circular region is used for the source and a co-located, disjointed annulus is used for the background. Generally, the background should be extracted from a region near the source. This is a good estimation of the background as long as the annulus does not intersect other sources, as it is independent on the instrument biases. This technique cannot be applied to extended sources, such as IC~5063. 

What is usually done for imagery of AGNs is to study the evolution of the flux along a virtual line that starts from the SMBH location and that extends in a direction perpendicular to winds and jets axis. When this flux reaches a plateau, the value of this plateau is assumed to be the background flux. This requires that the source is extended, but not to the point of reaching the borders of the FoV. 

In order to generalize this process for complex structures with potentially polluted background (such as the dust lane in IC~5063), we used a method that is not common in astrophysics \citep[see e.g.][]{Almoznino1993} but that is frequently used in astrophotography \citep{Bijaoui1980}. The method, as presented in Sect.~\ref{Pipeline:Error}, is based on the counts histogram of the stacked pictures. This mode calculation assumes the sky radiation is normally distributed around a typical sky value. Thus if the distribution of the image intensities is constructed, the sky pixels will practically form a Gaussian centered on the sky level. Other radiating sources in the frame, which represent additional photon contributions on top of the sky, will form a tail on the bright side of the sky distribution (see Fig.~\ref{fig:NGC_1068_error_histograms}). The advantage of this approach is that, no matter how bright these sources are, they will not affect the location of the peak of this histogram and hence the measured sky value. Other "bad" pixels, such as cosmic ray hits and defects of the CCD chip, that are either fainter or brighter than the sky will (for the most part) not affect the peak location. The sky value can be taken either as the exact location of the histogram peak (mode) or as the peak of a Gaussian or parabola fitted to the histogram in its peak region. To estimate the background noise, we constructed the distribution of the image intensities using a logarithmic scale and a statistic rule to bin the intensities. The background value is then chosen to be the peak of the obtained histogram (mode).

\section{Conclusion}
\label{Conclusion}
In this first paper of a series, we present a general pipeline for the reduction of UV and optical polarimetric data from the final re-calibrated archive of the HST/FOC legacy instrument. This code, written in \textsc{python} language, is designed to bring polarimetric analysis to the community. We checked and validated our tool by re-reducing NGC~1068's data with even a slightly better precision that what was done a few decades ago. Moreover, it allowed us to begin the analysis of the UV polarization map of IC~5063. We were able to use the full power of the space-resolved polarization to divide the analysis in regions of different characteristics. With this, the pipeline proves to be a powerful tool to make use of the still unmatched polarimetric capabilities of the FOC. Its ease and promptness of use will allow us to homogeneously reduce and analyse the full AGN sample among the FOC archive to get a better understanding of their overall properties and present them in the subsequent papers of the series. However, due to the fact that the FOC is the last mid- to far-UV polarimeter to have been in operation, this study can only be completed or taken further with future instruments. This study aim then at the preparation of future spectro-polarimeter instruments, such as POLLUX, planned to be mounted on the Habitable Wold Observatory by NASA.

\begin{acknowledgements}
The authors would like to acknowledge the referee for their comments that helped to improved this paper. TB and FM would also like to acknowledge the great help of Drs. Morganti and Maksym, who kindly shared their ACTA and HST maps of IC~5063, respectively. The authors would also like to deeply thank Dr. Antonucci for his comments and help through the pipeline creation and results analyses. 
\end{acknowledgements}

\bibliographystyle{aa} 
\bibliography{biblio}

\end{document}